\documentclass[aps,prd,twocolumn,showpacs,superscriptaddress,groupedaddress, nofootinbib]{revtex4} 

\usepackage{graphicx}
\usepackage{dcolumn}
\usepackage{bm}
\usepackage{color}

\usepackage[colorlinks=true]{hyperref}
\hypersetup{citecolor = blue}

\usepackage{multirow}
\usepackage{newtxtext,newtxmath}
\usepackage{mathrsfs}
\usepackage[utf8]{inputenc}
\usepackage{graphicx}                           
\usepackage{amsmath}	
\usepackage{pdflscape}	
\usepackage[T1]{fontenc}
\usepackage{ae,aecompl}
\usepackage{ulem}

\begin{document}


\title{Improved Cosmological Constraints from Morphology-Based Marked Correlation Functions}%

 \author{Xu Xiao}
 \affiliation{School of Physics and Astronomy, Sun Yat-Sen University, Zhuhai 519082, China}

\author{Zhao Chen}
\affiliation{Tsung-Dao Lee Institute, Shanghai Jiao Tong University, Shanghai 200240, China}
\affiliation{Department of Astronomy, School of Physics and Astronomy, Shanghai Jiao Tong University, Shanghai 200240, China}
\affiliation{State Key Laboratory of Dark Matter Physics, School of Physics and Astronomy, Shanghai Jiao Tong University, Shanghai 200240, China}
\affiliation{Key Laboratory for Particle Astrophysics and Cosmology (MOE)/Shanghai Key Laboratory for Particle Physics and Cosmology, Shanghai 200240, China}

\author{Yu Yu}
\affiliation{Department of Astronomy, School of Physics and Astronomy, Shanghai Jiao Tong University, Shanghai 200240, China}
\affiliation{State Key Laboratory of Dark Matter Physics, School of Physics and Astronomy, Shanghai Jiao Tong University, Shanghai 200240, China}
\affiliation{Key Laboratory for Particle Astrophysics and Cosmology (MOE)/Shanghai Key Laboratory for Particle Physics and Cosmology, Shanghai 200240, China}

 \author{Xiao-Dong Li}
  \email{lixiaod25@mail.sysu.edu.cn}
 \affiliation{School of Physics and Astronomy, Sun Yat-Sen University, Zhuhai 519082, China}
 \affiliation{Peng Cheng Laboratory, Shenzhen, Guangdong 518066, China}

\affiliation{CSST Science Center for the Guangdong–Hong Kong–Macau Greater Bay Area, SYSU, Zhuhai 519082, China}
 \author{Le Zhang}
 \email{zhangle7@mail.sysu.edu.cn}
 \affiliation{School of Physics and Astronomy, Sun Yat-Sen University, Zhuhai 519082, China}
 \affiliation{CSST Science Center for the Guangdong–Hong Kong–Macau Greater Bay Area, SYSU, Zhuhai 519082, China}
\date{\today}

\begin{abstract}
The cosmic web contains morphology-dependent information that is not fully captured by standard two-point statistics. We construct morphology-based marked correlation functions (MCFs) by assigning marks to halos according to the cosmic-web morphology identified with the \textsc{Nexus} algorithm. Using the \textsc{Kun} simulation suite, which spans 129 $w_0w_a$CDM cosmologies, we build Gaussian-process emulators for the MCFs as functions of cosmological parameters and tracer bias. We then apply the emulators to mock halo catalogues from the independent \textsc{Jiutian} simulation and perform a joint likelihood analysis to quantify the resulting cosmological constraints. We consider two marker choices: a discrete morphology marker and a continuous morphology strength marker. The continuous marker improves the Figure of Merit (FoM) by a factor of $\sim 8.6$ relative to the standard 2PCF and reduces the $1\sigma$ uncertainty on $\sigma_8$ by a factor of $\sim 5$. The discrete marker gives a more modest FoM improvement of $\sim 17\%$. We further test the impact of tracer selection by varying the halo mass threshold by a factor of $\sim 4.5$. Even for the lowest mass threshold, the continuous marker remains unbiased and achieves a FoM about $\sim 3.4$ times higher than that of the 2PCF alone. These results show that morphology-based MCFs, combined with simulation-based emulation, provide a useful framework for extracting additional cosmological information from large-scale structure surveys.
\end{abstract}

\maketitle

\section{Introduction}
\label{sec:introduction}

Over the past two decades, galaxy redshift surveys such as the 2dF Galaxy Redshift Survey (2dFGRS) \citep{2df:Colless:2003wz}, the 6dF Galaxy Survey (6dFGS) \citep{beutler20116df}, the WiggleZ Dark Energy Survey \citep{blake2011wigglez,blake2011wigglezb}, and the Sloan Digital Sky Survey (SDSS) \citep{york2000sloan,Eisenstein:2005su,Percival:2007yw,anderson2012clustering,alam2017clustering} have established large-scale structure as a central probe of cosmology. The current generation of Stage-IV large-scale-structure surveys, including the Dark Energy Spectroscopic Instrument (DESI) \citep{desi2016}, the Vera C. Rubin Observatory Legacy Survey of Space and Time (LSST) \citep{lsst2009}, the Euclid satellite \citep{eucild2011,eucild2024}, the Nancy Grace Roman Space Telescope \citep{rst2019}, and the Chinese Space Station Telescope (CSST) \citep{gong2019csst}, will map the matter distribution of the Universe with unprecedented volume and precision. This dramatic increase in data quality motivates the development of statistical tools capable of extracting cosmological information beyond the standard two-point description, especially from the non-linear and non-Gaussian cosmic web.

The two-point correlation function (2PCF) and its Fourier-space counterpart, the power spectrum, have long served as the standard summary statistics for large-scale-structure analyses \citep{Kaiser,Ballinger,Eisenstein_1998,Blake_2003,Seo_2003}. Their computational efficiency and direct sensitivity to the expansion history and the growth of structure make them indispensable for constraining cosmological models. These statistics have been successfully applied to a wide range of galaxy surveys \citep{2dFGRS,6dFGRS,WiggleZ2011B,WiggleZ2011c,SDSS_York,Eisenstein:2005su,Percival:2007yw,anderson2012clustering,sanchez2012clustering,sanchez2013clustering,anderson2014clustering,samushia2014clustering,ross2015clustering,beutler2016clustering,sanchez2016clustering,alam2017clustering,chuang2017clustering}. However, gravitational instability generates non-Gaussian structures on non-linear scales, producing a complex cosmic web composed of clusters, filaments, walls, and voids. While the 2PCF provides a complete statistical description of a Gaussian random field, it cannot fully characterize the morphology-dependent information encoded in the evolved cosmic web. As a result, cosmological information associated with non-Gaussian structure formation may remain inaccessible to standard two-point analyses.

Several classes of statistics have been developed to access this additional information. These include three-point \citep{Sabiu2016,Slepian_2017} and four-point correlation functions \citep{Sabiu_2019}, void statistics \citep{ryden1995measuring,lavaux2012precision}, topological summaries, and machine-learning-based approaches \citep{Ravanbakhsh17,Mathuriya18,pan2020cosmological}. These methods have demonstrated the importance of non-Gaussian information for cosmological inference. At the same time, they often face practical challenges related to computational cost, covariance estimation, interpretability, or the construction of fast theoretical predictions. This motivates complementary summary statistics that remain physically interpretable while capturing information beyond the standard 2PCF.

One physically motivated route is to exploit the morphology of the cosmic web itself. Cosmic-web classification divides the matter distribution into morphologically distinct components, commonly identified as clusters, filaments, walls, and voids. Early studies based on N-body simulations showed that these morphology classes exhibit different halo mass functions and halo occupation properties \citep{Hahn_2007,Miguel_2010}, evolve significantly over cosmic time \citep{Hahn_2007}, and possess characteristic sizes and internal structures \citep{Miguel_2010,Bond_2010}. These results have motivated the development of robust cosmic-web identification algorithms. Among them, Hessian-based methods applied to the density, tidal, or velocity-shear fields have proven particularly effective \citep{Arag_n_Calvo_2007,Wu_2009,Forero_Romero_2009,Bond_2010,Cautun_2012}. Such methods identify coherent structures by extracting the local morphological signature of the underlying field across multiple spatial scales, thereby providing a physically interpretable segmentation of the cosmic web.

Cosmic-web morphology has subsequently been used to study a wide range of structure-formation effects, including the morphology dependence of halo properties \citep{Hellwing_2021}, the power spectra of different cosmic-web components in real and redshift space \citep{Bonnaire_2022,Bonnaire_2023,Sunseri_2025}, the impact of massive neutrinos on different morphology classes \citep{Khoshtinat_2024}, the theoretical volume and mass fractions of the cosmic web \citep{Ay_oberry_2024,Dome_2023}, and the influence of baryonic feedback on the matter distribution within the cosmic web \citep{dong_2025}. Cosmic-web morphology has also been explored as a cosmological probe in its own right. Examples include the cosmological sensitivity of void properties \citep{Lee_2009,Biswas_2010}, critical-point counts \citep{Codis_2013}, and topological descriptors such as Betti numbers \citep{Feldbrugge_2019}. These studies indicate that the morphology of the cosmic web contains cosmological information that is not fully captured by conventional two-point statistics.

Although these studies demonstrate the cosmological relevance of cosmic-web morphology, incorporating this information into precision inference remains non-trivial. A direct analysis of clusters, filaments, walls, and voids would require modeling multiple morphology-dependent fields and their auto- and cross-correlations, leading to a high-dimensional data vector and a complicated covariance structure. Moreover, cosmic-web classification is usually defined from a smoothed density, tidal, or velocity-shear field, whereas galaxy surveys provide sparse and biased tracers of the underlying matter distribution. A practical approach should therefore compress morphology-dependent information into a statistic that remains close to standard clustering measurements and can be calibrated for biased tracers.

The marked correlation function (MCF) provides such a framework \citep{Beisbart:2000ja,Beisbart2002,Gottl2002,Sheth:2004vb,Sheth:2005aj,Skibba2006,White_2009,White2016,Satpathy:2019nvo,massara2020,Philcox2020}. The MCF extends the standard 2PCF by assigning each tracer a weight, or mark, determined by a chosen physical property. It then measures the clustering of marked tracers relative to the unmarked distribution, with the standard 2PCF recovered when all marks are equal. By choosing marks that depend on local density, galaxy properties, or cosmic-web morphology, the MCF can enhance the contribution of specific physical regimes while retaining the simplicity of a two-point statistic. Previous applications to mock catalogues \citep{MCF_Yang,Xiao2022MCF,xiao2026} and SDSS data \citep{Lai2024MCF} have shown that density-dependent marks can extract cosmological information beyond the standard 2PCF and improve parameter constraints.

In this work, we construct a morphology-based MCFs by assigning marks according to cosmic-web morphology identified with the \textsc{Nexus} algorithm. We consider two marking schemes: a discrete morphology-class mark and a continuous morphological-strength mark. The former weights tracers according to their \textsc{Nexus}-identified morphology class, while the latter retains information about the strength of the local cosmic-web signature. This construction embeds cosmic-web morphology into a compact two-point statistic, linking physically interpretable structure classification with standard clustering analyses.

Because the morphology-weighted MCF depends on the non-linear evolution of the cosmic web, a closed-form analytic prediction is difficult to obtain. We therefore adopt a simulation-based emulation approach \citep{Cranmer_2020}. Using the \textsc{Kun} simulation suite, which spans 129 $w_0w_a$CDM cosmologies, we train Gaussian process emulators for the morphology-dependent MCF as functions of cosmology and tracer bias. We validate the emulator against the independent \textsc{Jiutian} simulation suite and perform a joint likelihood analysis to quantify the resulting cosmological constraints. This approach builds on the broader development of cosmological emulators, from early emulators for the matter power spectrum \citep{2009ApJ...705..156H,2010ApJ...715..104H,2010ApJ...713.1322L} to more recent extensions targeting non-linear scales and alternative summary statistics \citep{2014ApJ...780..111H,2017ApJ...847...50L,bocquet2020miratitan,Kwan2015,Wibking_2019,Kwan_2013,Nishimichi_2019,kobayashi2020accurate,Kwan2023,Moran2023}.

Our results show that morphology-weighted MCFs provide a practical route to extracting cosmological information from the non-linear cosmic web. By combining \textsc{Nexus}-based morphology classification with simulation-based emulation, we obtain a statistic that is physically interpretable, computationally tractable, and more informative than the standard 2PCF alone. This framework offers a promising pathway for incorporating cosmic-web morphology into cosmological analyses of current and upcoming galaxy redshift surveys.

The structure of this paper is as follows. In Section \ref{sec:data}, we describe the suite of N-body simulations used to train the emulator and to construct validation datasets and covariance estimates. Section \ref{sec:methodology} presents our methodology, including the mock catalogs, the \textsc{Nexus} cosmic web identification, the construction of MCFs, the Gaussian process regression framework for emulator development, and the covariance estimation. In Section \ref{sec:results}, we present our main results, demonstrating the constraining power of the Morphology-based MCF relative to the standard 2PCF, and exploring its dependence on scale, mark definition, and halo mass bias. We conclude with a summary and discussion of future prospects in Section \ref{sec:conclusion}.

\section{Data}
\label{sec:data}

\subsection{The \textsc{Kun} simulation}
\label{subsec:kun_simulation}

In this work, we use the \textsc{Kun} simulation suite as the training set for the Gaussian process emulator~\citep{yu2025kunsimulation}. The \textsc{Kun} suite is designed to sample a broad cosmological parameter space suitable for simulation-based inference in dynamical dark-energy models. It consists of 129 cosmologies in an extended $w_0w_a{\rm CDM}+\sum m_\nu$ parameter space, including one fiducial cosmology based on the \textit{Planck} 2018 results~\citep{planck2018} and 128 non-fiducial cosmologies generated using a Sobol sequence~\citep{sobol1967distribution}. The Sobol design provides a quasi-random, space-filling sampling of the high-dimensional parameter space, which is well suited for emulator construction because it reduces large interpolation gaps within the training domain.

For each cosmology, the matter distribution is evolved using the \textsc{Gadget-4} $N$-body solver\footnote{https://wwwmpa.mpa-garching.mpg.de/gadget4/}~\citep{Springel2021gadget}. Each simulation adopts a periodic cubic box with side length $1~h^{-1}{\rm Gpc}$ and contains $3072^3$ particles, corresponding to a particle mass of $2.87\,(\Omega_{m}/0.3)\times10^9~h^{-1}M_{\odot}$. This volume and mass resolution allow the simulations to resolve the halo population relevant for the tracer samples used in this analysis while maintaining a sufficiently large volume for robust measurements of clustering statistics.

The cosmological parameter space spans baryon density $\Omega_b$, matter density $\Omega_{m}$, scalar spectral index $n_s$, Hubble constant $H_0$, primordial fluctuation amplitude $A_s$, dark energy equation-of-state parameters $(w_0, w_a)$, and the summed neutrino mass $\sum m_\nu$, with ranges
$\Omega_b \in [0.04, 0.06]$,
$\Omega_{m} \in [0.24, 0.40]$,
$n_s \in [0.92, 1.00]$,
$H_0 \in [60, 80]~{\rm km\,s^{-1}\,Mpc^{-1}}$,
$A_s \in [1.70, 2.50]\times10^{-9}$,
$w_0 \in [-1.30, -0.70]$,
$w_a \in [-0.50, 0.50]$, and
$\sum m_{\nu} \in [0.00, 0.30]~{\rm eV}$.

\begin{figure}[htpb]
\centering
\includegraphics[scale=0.49]{"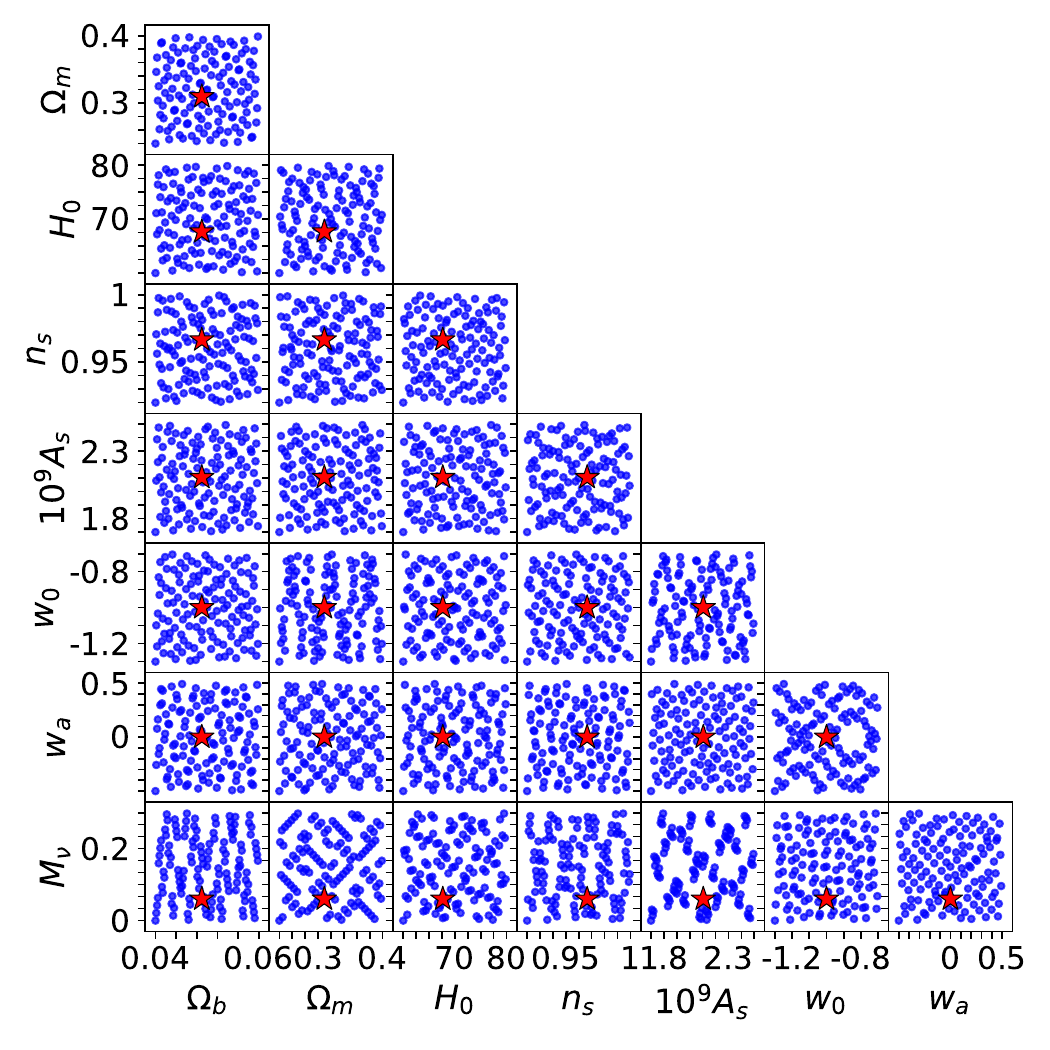"}
\caption{Sampling of the 129 \textsc{Kun} cosmologies used to train the emulator. Blue points show the 128 Sobol-sequence cosmologies, and the red star marks the fiducial \textit{Planck} 2018 model. The broad parameter coverage ensures that the emulator operates primarily in the interpolation regime, thereby improving accuracy in the likelihood analysis.}
\label{fig:param_space}
\end{figure}

Figure~\ref{fig:param_space} shows the distribution of the cosmological design points. The blue points denote the 128 Sobol-sampled cosmologies, while the red star marks the fiducial \textit{Planck} 2018 cosmology. The approximately uniform coverage of the design space helps improve the interpolation accuracy of the emulator and limits the need for extrapolation in the subsequent likelihood analysis.

Each cosmology is realized once. To reduce the impact of sample variance from large-scale modes, the initial conditions adopt the fixed-amplitude method~\citep{2016MNRAS.462L...1A}. Halo and subhalo catalogues are identified using the \textsc{Rockstar} phase-space halo finder~\citep{Behroozi2012rockstar}. These catalogues are used to construct the halo tracer samples on which the morphology-dependent marked correlation functions are measured.

The \textsc{Kun} suite is part of the broader \textsc{Jiutian} simulation program developed for CSST-related cosmological applications~\citep{han2025jiutiansimulationscsstextragalactic,gong2019csst}. Other simulations in the program cover different resolutions, cosmological models, neutrino scenarios, dark-matter and dark-energy models, zoom-in regions, and mock galaxy catalogue constructions~\citep{universe11070212,Yu_2026,2023MNRAS.526.3156H,2019ApJ...875L..11Z,2024ApJ...966..236L,2024MNRAS.529.4958P,2024MNRAS.529.4015G,tan2025semianalyticalmockgalaxycatalog,wei2025mockobservationscsstmission,Wei_2026}. In the present analysis, however, the emulator training is based specifically on the 129 cosmologies of the \textsc{Kun} suite.

\subsection{The \textsc{Jiutian} simulation}
\label{subsec:jiutian_simulation}

We use the primary \textsc{Jiutian} simulation as an independent dataset for emulator validation and covariance estimation~\citep{han2025jiutiansimulationscsstextragalactic}. Unlike the \textsc{Kun} suite, which samples a broad cosmological parameter space for emulator training, the primary \textsc{Jiutian} simulation adopts a single fiducial \textit{Planck} 2018 $\Lambda$CDM cosmology~\citep{planck2018}. The cosmological parameters are $\Omega_m=0.3111$, $\Omega_\Lambda=0.6889$, $\Omega_b=0.049$, $n_s=0.9665$, $\sigma_8=0.8102$, and $H_0=67.66~{\rm km~s^{-1}~Mpc^{-1}}$. The simulation evolves $6144^3$ particles in a periodic box with side length $2~h^{-1}{\rm Gpc}$. Halo and subhalo catalogues are identified using the friends-of-friends algorithm and \textsc{Subfind}~\citep{fof,Springel_2001subfind}.

The large simulation volume of \textsc{Jiutian} makes it well suited for validating the emulator predictions at the fiducial cosmology and for estimating the covariance matrix of the measured statistics. Since \textsc{Jiutian} is independent of the \textsc{Kun} training simulations, it provides a useful test of whether the emulator can accurately predict the morphology-dependent marked correlation functions for a simulation that is not included in the training set.

For the fiducial tracer sample, we impose a fixed number density,
$\bar n=10^{-3}~(h^{-1}{\rm Mpc})^{-3}$, by ranking halos and subhalos by mass and selecting the most massive objects. This selection corresponds to a characteristic mass threshold of approximately $M_{\rm cut}\simeq3.85\times10^{12}~h^{-1}M_\odot$ in both the \textsc{Kun} and \textsc{Jiutian} simulations. Fixing the tracer number density reduces the impact of abundance differences across simulations, so that variations in the measured correlation functions primarily reflect changes in clustering and cosmic-web morphology. The adopted number density is also comparable to that of current and forthcoming spectroscopic galaxy surveys.

To study the dependence of our method on tracer bias, we additionally construct halo samples using three different mass thresholds: $M_{\mathrm{cut}} = \{3.85 \times 10^{12}, 2.45 \times 10^{12}, 0.85 \times 10^{12}\}~h^{-1}M_{\odot}$.
For the \textsc{Kun} suite, these selections are applied to all 129 cosmologies, yielding a total of $129\times3=387$ halo samples. For the primary \textsc{Jiutian} simulation, we apply the same three mass thresholds to generate the corresponding fiducial and bias-varied validation samples. These samples allow us to test whether the morphology-dependent MCF remains stable when the tracer population changes substantially.

We characterize the large-scale tracer bias using the average ratio between the halo power spectrum and the linear matter power spectrum. Specifically, we define
\begin{equation}
b^2 = \frac{1}{k_{\rm max}}\int_{0}^{k_{\max}}
\frac{P_{\rm halo}(k)}{P_{m}(k)}\,{\rm d}k\,,
\label{eq:bias}
\end{equation}
where $P_{\rm halo}(k)$ is the halo power spectrum measured from the simulation and $P_{m}(k)$ is the corresponding linear matter power spectrum computed with \textsc{CAMB}~\citep{Lewis_2000}. We adopt
$k_{\max}=0.1~h{\rm Mpc}^{-1}$,
so that the estimate is restricted to large scales where the tracer bias is expected to be approximately scale independent.

\begin{figure}[htpb]
	\centering
	\includegraphics[scale=0.5]{"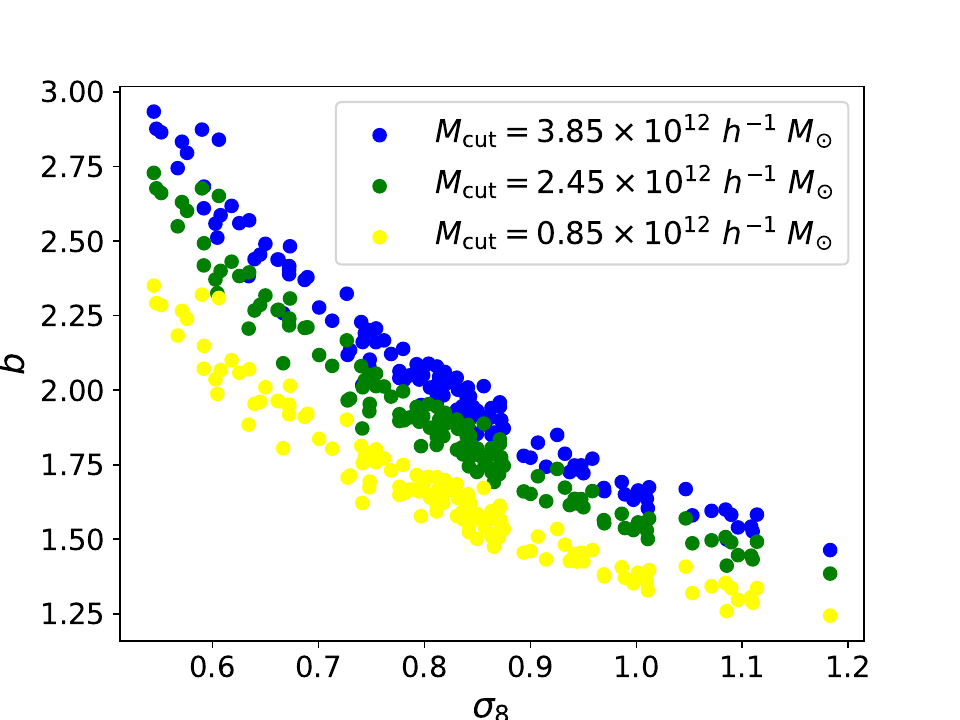"}
    \caption{The joint distribution of the tracer bias $b$ and the cosmological parameter $\sigma_8$ for different mass thresholds $M_{\mathrm{cut}}$ from the Kun simulations. The blue, green, and yellow points correspond to $M_{\mathrm{cut}} = 3.85 \times 10^{12}, 2.45 \times 10^{12},$ and $0.85 \times 10^{12}\, h^{-1}M_{\odot}$, respectively.}
\label{fig:hist2d_bias_sigma8}
\end{figure}

Figure~\ref{fig:hist2d_bias_sigma8} shows a negative correlation between the large-scale tracer bias $b$ and $\sigma_8$ for all three mass-threshold samples. This trend is expected for a fixed-$M_{\rm cut}$ selection: when $\sigma_8$ is larger, halos above a given mass threshold correspond to less rare density peaks and therefore have a lower clustering bias. The bias amplitude also decreases for lower $M_{\rm cut}$, reflecting the weaker clustering of less massive halos. This correlation illustrates that tracer bias is coupled to both cosmology and sample selection. We therefore compare different $M_{\rm cut}$ samples to test the robustness of the morphology-dependent marked correlation functions against tracer-bias variations.

Since the clustering statistics used in this work are intended to mimic measurements from spectroscopic galaxy surveys, we measure them in redshift space. After constructing the halo tracer samples, we displace each object along the line of sight according to its peculiar velocity. Redshift-space positions are constructed under the distant-observer approximation, taking the simulation $z$-axis as the line of sight:
\begin{equation}
\bm{s} = \bm{r} + \frac{\bm{v}\cdot\hat{\bm{z}}}{aH(a)}\,\hat{\bm{z}},
\label{eq:rsd}
\end{equation}
where $\bm{r}$ and $\bm{s}$ denote the real- and redshift-space positions, $\bm{v}$ is the peculiar velocity, $a$ is the scale 
factor, and $H(a)$ is the Hubble parameter.

\section{Methodology}
\label{sec:methodology}

\subsection{Cosmic web classification}\label{sec:method}

We classify the simulation volume into four types of cosmic environments: knots, filaments, walls, and voids. Here, knots refer to the densest node-like regions, filaments are elongated bridges between knots, walls are sheet-like structures, and voids are underdense regions left after the other structures are identified. Our classification follows the \textsc{Nexus} method \citep{Cautun_2012}, which is based on the local shape of the smoothed density field.

The procedure can be understood in three steps. First, the density field is smoothed on several spatial scales. Second, the local shape of the field is measured from the Hessian matrix. Third, the strongest structure signal across all smoothing scales is used to assign each cell to one cosmic environment.

\subsubsection{Density maps on different scales}

We start from the density contrast field $\delta(\mathbf{x})$ and use its logarithmic form to reduce the large contrast between high-density and low-density regions. For each smoothing scale $R_i$, the logarithmic density field is convolved with a Gaussian filter,
\begin{equation}
f_{R_i}(\mathbf{x}) = G_{R_i}\left[\log_{10}\left(\delta(\mathbf{x})+1\right)\right].
\end{equation}
Using several smoothing scales is important because cosmic structures do not have a single characteristic size. Small smoothing scales are more sensitive to compact objects and narrow filaments, while larger smoothing scales trace extended walls and large-scale connections.

For each smoothed density map, we compute the Hessian matrix,
\begin{equation}
\mathbf{H}_{ij,R_i}(\mathbf{x}) =
R_i^2
\frac{\partial^2 10^{f_{R_i}(\mathbf{x})}}
{\partial x_i \partial x_j}.
\end{equation}
The factor $R_i^2$ is included to make the response from different smoothing scales more comparable. The three eigenvalues of the Hessian matrix, ordered as
$\lambda_1 \leq \lambda_2 \leq \lambda_3$, describe how the density field bends along three principal directions. Therefore, their signs and relative amplitudes provide a simple way to describe the local shape of the density field.

In this Hessian-based picture, a knot is a region where the density field curves inward along all three directions. A filament curves inward along two directions but remains extended along the third. A wall curves inward mainly along one direction and is extended in the other two directions. These ideas correspond to the following initial conditions: knots have $\lambda_1,\lambda_2,\lambda_3 < 0$, filaments have $\lambda_1,\lambda_2 < 0$, and walls have $\lambda_1 < 0$.

\subsubsection{Structure strength parameters}

The sign of the eigenvalues gives a first indication of the local environment, but it is not enough to produce a clean classification. A cell may satisfy more than one condition at the same time, and weak fluctuations can also pass the sign test. Therefore, \textsc{Nexus} assigns a structure strength to each environment type. This strength measures how clearly a cell behaves like a knot, filament, or wall at a given smoothing scale.

Following \citet{Cautun_2012}, the scale-dependent {\it strength parameters} are defined  as
\begin{align}
S^{k}_R(\mathbf{x}) &= \frac{\lambda_3^2}{|\lambda_1|}\,,\\
S^{f}_R(\mathbf{x}) &= \frac{\lambda_2^2}{|\lambda_1|}
\left(1 - \left|\frac{\lambda_3}{\lambda_1}\right|\right)\,,\\
S^{w}_R(\mathbf{x}) &= |\lambda_1|
\left(1 - \left|\frac{\lambda_2}{\lambda_1}\right|\right)
\left(1 - \left|\frac{\lambda_3}{\lambda_1}\right|\right)\,.
\end{align}
Here, $S^\alpha$ with $\alpha \in \{k,f,w\}$  denote the strengths of knots, filaments, and walls, respectively. A larger value means that the local density field more clearly resembles the corresponding structure type.

Since the same physical structure may be most clearly seen at different smoothing scales, we keep the maximum response over all scales,
\begin{equation}
S^\alpha(\mathbf{x}) = \max_{R_i} S^\alpha_{R_i}(\mathbf{x})\,.
\label{eq:strength_max}
\end{equation}
This step combines information from different scales into one final strength field for a given type of structure. In practice, we use a set of smoothing scales, which is chosen to be comparable to the grid spacing of the density field, while the largest scale is chosen to cover the typical size of prominent cosmic web structures. Following ~\citet{Sunseri_2025}, we adopt large-scale smoothing set $R_{\rm large} = R_0 \times \{8, 8\sqrt{2}, 16\}$, where $R_0 \approx 1.95~h^{-1}{\rm Mpc}$. We also adopt small-scale smoothing set $R_{\rm small} = R_0 \times \{4, 4\sqrt{2}\}$ for comparison.

\subsubsection{Morphology Markers for Halo Samples}

After the cosmic-web morphology is identified by \textsc{Nexus}, each halo or subhalo can be assigned a morphology marker according to its position in the density field. For a halo at $\mathbf{x}_i$, we denote its morphology marker as $m_i^\alpha$, where $\alpha \in \{k,f,w\}$ corresponds to knots, filaments, and walls. Void regions are not considered in this work. We consider two possible definitions of the marker.  The first one is a {\it discrete marker} based on the \textsc{Nexus} morphology classification, while the second one is a {\it continuous marker} based on the local structure strength:
\begin{equation}
m_i^\alpha =
\begin{cases}
1, & c_i=\alpha\,,\\
0, & c_i\neq \alpha\,,
\end{cases}
\qquad \mathrm{or} \qquad
m_i^\alpha = S^\alpha(\mathbf{x}_i)\,.
\label{eq:morphology_marker}
\end{equation}
Here, $c_i$ is the discrete morphology class assigned to the halo or subhalo, and $S^\alpha(\mathbf{x}_i)$ is the \textsc{Nexus} strength of morphology $\alpha$ sampled at the halo position. Therefore, the same notation $m_i^\alpha$ can describe either a hard morphology classification or a continuous strength-based weighting.

For the discrete-marker case, the halo sample is explicitly divided into different morphology-selected subsamples. We denote the overdensity field of halos with morphology $\alpha$ by $\delta^\alpha(\mathbf{x})$. The total halo overdensity field can then be formally decomposed as $\delta(\mathbf{x}) = \sum_{\alpha}\delta^\alpha(\mathbf{x})$. In this case, MCF measures the autocorrelation of halos or subhalos belonging to the same morphology:
\begin{equation}
\begin{aligned}
W^\alpha_{\rm dis}(\mathbf{r})=
\left\langle 
\delta^\alpha(\mathbf{x}) \delta^\alpha(\mathbf{x}+\mathbf{r}) \right\rangle \,.
\end{aligned}
\label{eq:mcf_discrete}
\end{equation}
Since $m^\alpha$ is a binary selector in the discrete case, this is equivalent to computing 2PCF of the morphology-selected halo sample. For example, $W^k_{\rm dis}$ describes the clustering of knot halos with other knot halos. 

For the continuous-marker case, we do not split the halo density field into different morphology subsamples. Instead, all halos or subhalos are kept in the same overdensity field $\delta(\mathbf{x})$, and the morphology information enters only through the continuous strength marker. MCF is therefore written as
\begin{equation}
W^\alpha_{\rm con}(\mathbf{r})=
\left\langle
\delta(\mathbf{x})m^\alpha(\mathbf{x})
\delta(\mathbf{x}+\mathbf{r})m^\alpha(\mathbf{x}+\mathbf{r})
\right\rangle \,.
\label{eq:mcf_continuous}
\end{equation}
Here, $m^\alpha(\mathbf{x})$ is given by the strength field $S^\alpha(\mathbf{x})$. This definition keeps the full halo sample fixed and weights each halo pair by the product of their local morphology strengths. Therefore, the discrete marker describes the clustering of halos selected by the same morphology class, whereas the continuous marker describes the clustering of the full halo sample weighted by its morphology strength.

\subsubsection{Scale Dependence of Morphology}

\begin{figure}[htpb]
    \centering
    \includegraphics[width=0.45\textwidth]{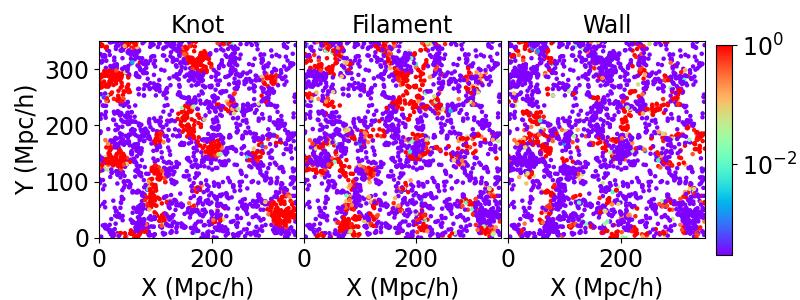}
    \includegraphics[width=0.45\textwidth]{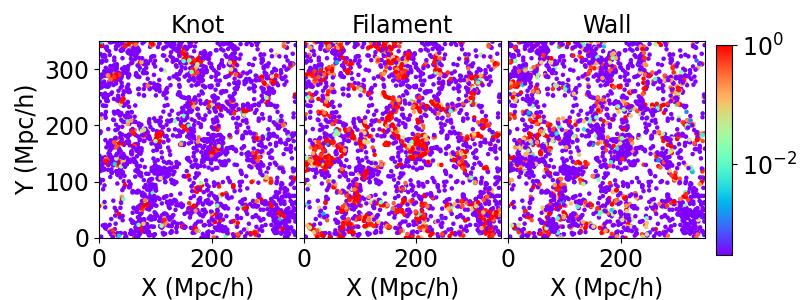}
    \caption{A slice of halos selected from the \textsc{Kun} simulation, colored by their discrete morphology classification: knot, filament, and wall. The top panels show the result obtained with the large-scale smoothing set $R_{\rm large} $. The bottom panels show the result obtained with the small-scale smoothing set
$R_{\rm small}$.}
    \label{fig:tag_comparison}
\end{figure}

\begin{figure}[htpb]
    \centering
    \includegraphics[width=0.45\textwidth]{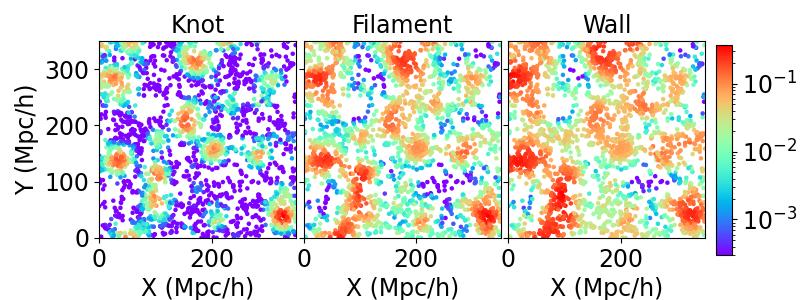}
    \includegraphics[width=0.45\textwidth]{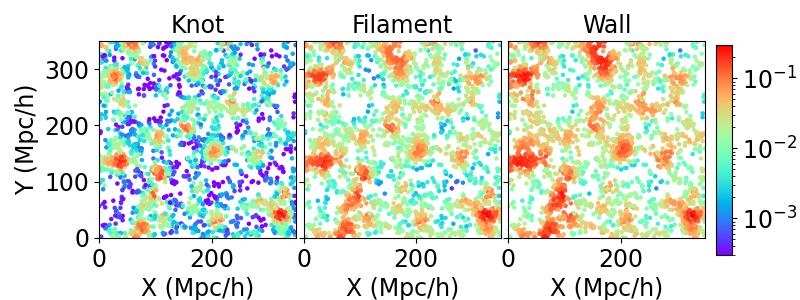}
    \caption{Same as Figure~\ref{fig:tag_comparison}, but for the continuous morphology strength markers. From left to right, the panels show the knot strengths $S^{k}(\mathbf{x})$, filament strengths $S^{f}(\mathbf{x})$, and wall strengths $S^{w}(\mathbf{x})$. The top panels correspond to the large-scale smoothing set $R_{\rm large} $, and the bottom panels correspond to the small-scale smoothing set $R_{\rm small} $. Unlike the discrete classification, the strength marker varies continuously and therefore keeps more information about the local morphology.}
    \label{fig:strength_comparison}
\end{figure}

\begin{figure}[htpb]
    \centering
    \includegraphics[width=0.45\textwidth]{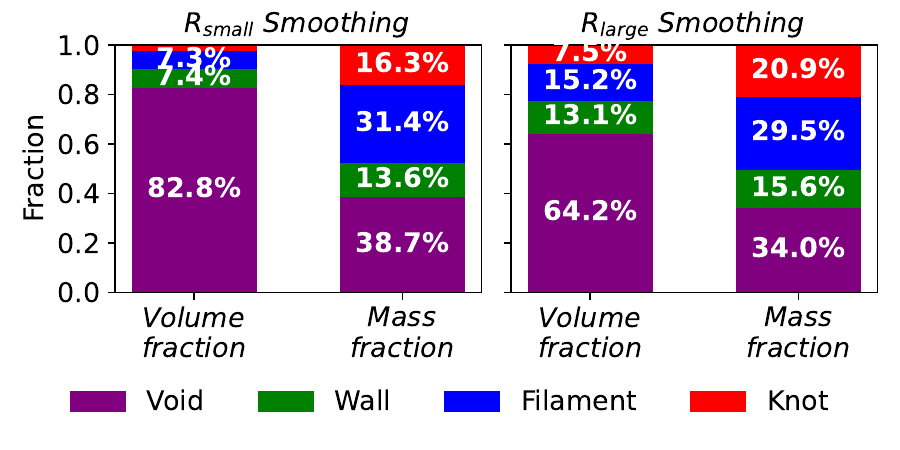}
    \caption{Mass and volume fractions of different cosmic-web morphologies, including knots, filaments, walls, and voids, measured from the \textsc{Jiutian} simulation. The results are shown for the two smoothing-scale choices used in this work. The two choices give somewhat different fractions, but both remain within the typical scatter found among different cosmic-web classification methods.}
    \label{fig:fractions}
\end{figure}

Figures~\ref{fig:tag_comparison} and~\ref{fig:strength_comparison} show how the morphology assignment depends on the smoothing scale. With the large-scale smoothing set, the identified structures are smoother and more connected. Knots, filaments, and walls form a coherent cosmic-web pattern. With the small-scale smoothing set, the classification becomes more fragmented, and the knot regions are much less common. This is expected because smaller smoothing scales are more sensitive to local fluctuations and discreteness effects in the halo distribution. The continuous strength markers show the same general trend. At large smoothing scales, the strength fields vary more smoothly across the halo distribution. At small smoothing scales, the strengths become more localized and fluctuate more strongly from halo to halo. Compared with the discrete tags, the continuous markers provide a softer description of morphology: a halo is not only assigned to one class, but can also carry information about how strongly it is associated with a knot, filament, or wall.

Figure~\ref{fig:fractions} gives a quantitative check of these morphology assignments. The mass and volume fractions change when we change the smoothing scale, especially for the knot component. However, such changes are expected because cosmic-web classification is known to depend on the details of the method, the smoothing scale, and the tracer sample. In our case, the halo density field is also affected by the adopted mass cut and the resulting tracer number density. Therefore, small differences from previous works, such as \citep{Sunseri_2025}, are not surprising.

For the main analysis, we use the large-scale smoothing set $R_{\rm large}$ as the fiducial choice. This choice gives a cleaner and more stable morphology classification. It is also less affected by small-scale discreteness and nonlinear details, which are harder to model accurately in a gravity-only simulation. The large-scale morphology therefore provides a more robust description of the density field for our goal of simulation-based cosmological inference. We still compare the two smoothing choices in the following sections, but we focus mainly on the large-scale result.

\subsection{Marked correlation functions}

In this work, the mark is chosen to describe the cosmic-web morphology of each halo. For the discrete morphology marker, the halo sample is first divided into knot, filament, and wall subsamples. The corresponding MCF is therefore computed using halos with the same morphology, as defined in Equation~\ref{eq:mcf_discrete}. In this case, $W^\alpha_{\rm dis}$ is essentially 2PCF of the morphology-selected halo sample, where $\alpha \in \{k,f,w\}$.

For the continuous morphology marker, we do not split the halo sample by morphology. Instead, all non-void halos are kept in the same density field, and the \textsc{Nexus} strength is used as a continuous weight. The corresponding statistic is given by Equation~\ref{eq:mcf_continuous}. In this case, $W^\alpha_{\rm con}$ measures how strongly the clustering of the full halo sample is associated with the local knot, filament, or wall strength.  By comparing these two weighting schemes, we aim to assess whether the discrete classification or the continuous case carries more discriminating power for cosmological inference.


In practice, we estimate the MCF using a weighted generalization of the Landy-Szalay estimator~\citep{Landy19932PCF}:
\begin{equation}
W(s, \mu) = \frac{WW - 2WR + RR}{RR},
\end{equation}
where $WW$, $WR$, and $RR$ denote the sums of the product of marks for galaxy-galaxy, galaxy-random, and random-random pairs, respectively, normalized by the total number of weighted pairs. Here, $s$ is the comoving pair separation, and $\mu \equiv \cos\theta$, with $\theta$ being the angle between the line of sight and the pair separation vector. When all marks are set to unity, the weighted pair counts reduce to the usual unweighted pair counts, and $W(s, \mu)$  becomes the standard Landy--Szalay estimator for 2PCF.

\begin{figure*}[htpb]
	\centering
		\includegraphics[width=0.9\textwidth] {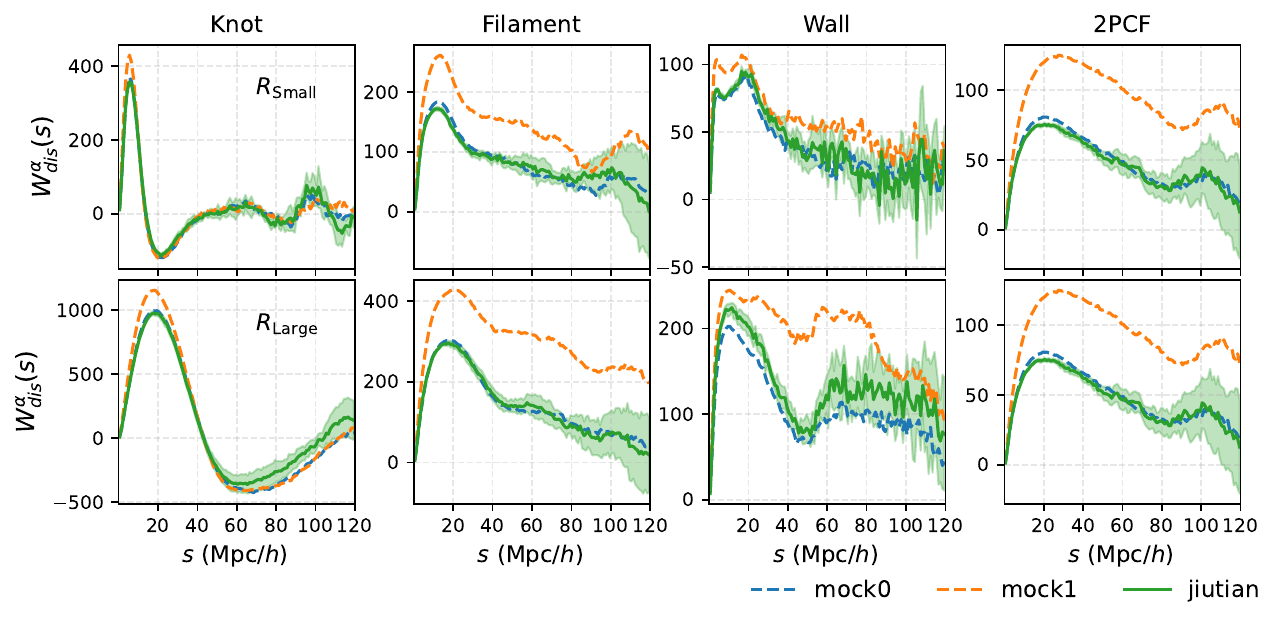}  \\
		\includegraphics[width=0.9\textwidth] {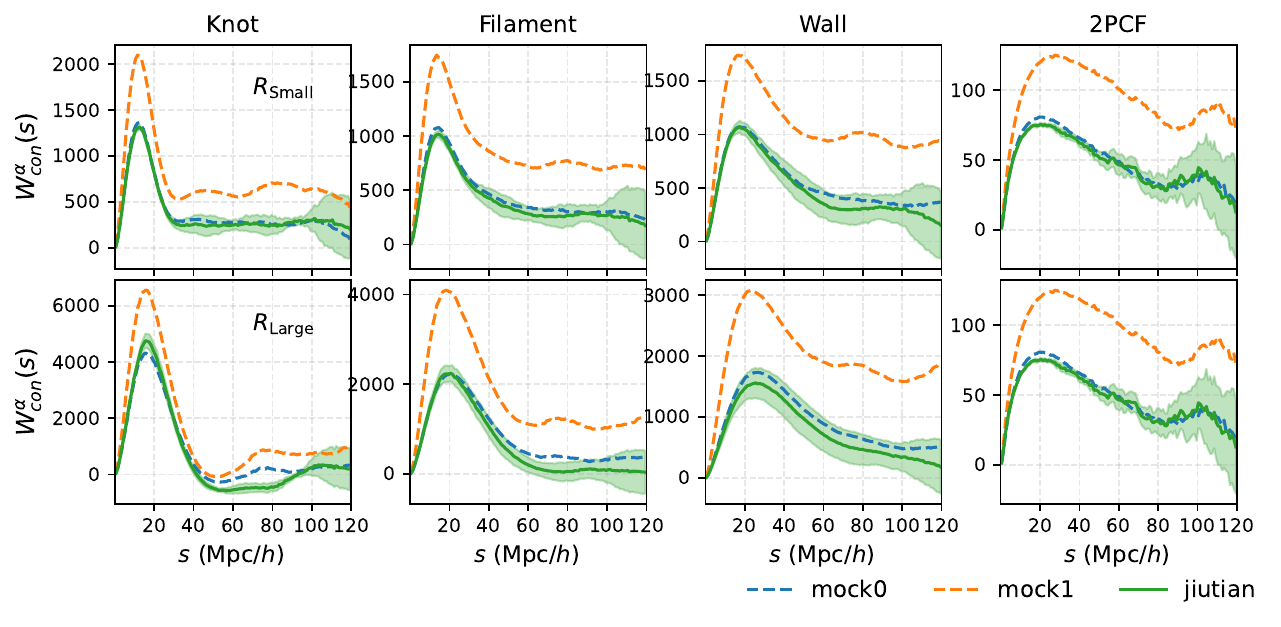}\\
\caption{Projected marked correlation functions, $s^2 W(s)$, at $z=0.5$. The upper and lower panels show the results for the discrete morphology marker and the continuous morphology strength marker, respectively.  Different coloured lines represent different simulation samples, including the fiducial cosmology from the \textsc{Jiutian} simulation, \textsc{Kun}  mock0, and the alternative cosmology from \textsc{Kun} mock1. Each sub-panel compares different smoothing-scale choices and different morphology types, including knots, filaments, and walls.}
\label{fig:xis_2PCF}
\end{figure*}

To construct a one-dimensional statistic with improved signal-to-noise ratio, we project the estimator $W(s,\mu)$ onto $\mu$ by integrating over the line-of-sight angle:
\begin{equation}
W(s) = \int_{\mu_{\min}}^{\mu_{\max}} W(s, \mu) \, d\mu,
\label{eq:intMs}
\end{equation}
where $\mu_{\min}=0$ and $\mu_{\max}=0.8$. The upper limit removes pairs that are nearly parallel to the line of sight, where redshift-space distortions, especially the Fingers-of-God effect, are strongest. We focus on the separation range
$15~h^{-1}{\rm Mpc} \leq s \leq 75~h^{-1}{\rm Mpc}$.
Figure~\ref{fig:xis_2PCF} shows the resulting projected MCFs at redshift $z=0.5$ for both the discrete and continuous morphology markers. 

As shown in Figure~\ref{fig:xis_2PCF}, the MCFs obtained from the discrete and continuous morphology markers show different behaviours. For the discrete marker, the halo sample is divided into knot, filament, and wall subsamples. Therefore, the measured MCFs correspond to the autocorrelations of halos with the same morphology. Their amplitudes and shapes differ significantly among the three morphology types, indicating that knots, filaments, and walls trace different clustering patterns. For the continuous strength marker, the full halo sample is kept fixed, and the morphology information only enters as a weight. In this case, the different MCFs mainly reflect how the underlying halo clustering is modulated by the knot, filament, or wall strength. As a result, the curves for different morphology strengths have more similar overall shapes than in the discrete-marker case.

In the $R_{\rm large}$ smoothing scheme, the characteristic clustering feature shifts to larger separations when moving from the knot strength $S^k$ to the filament strength $S^f$ and then to the wall strength $S^w$. This trend is physically reasonable: knots are the most compact structures, filaments extend over larger scales, and walls correspond to even broader sheet-like structures. The ordering of the MCF features therefore reflects the typical spatial scales of the different cosmic-web morphologies.

\begin{figure*}[htpb]
	\centering
    \includegraphics[width=0.47\textwidth]{"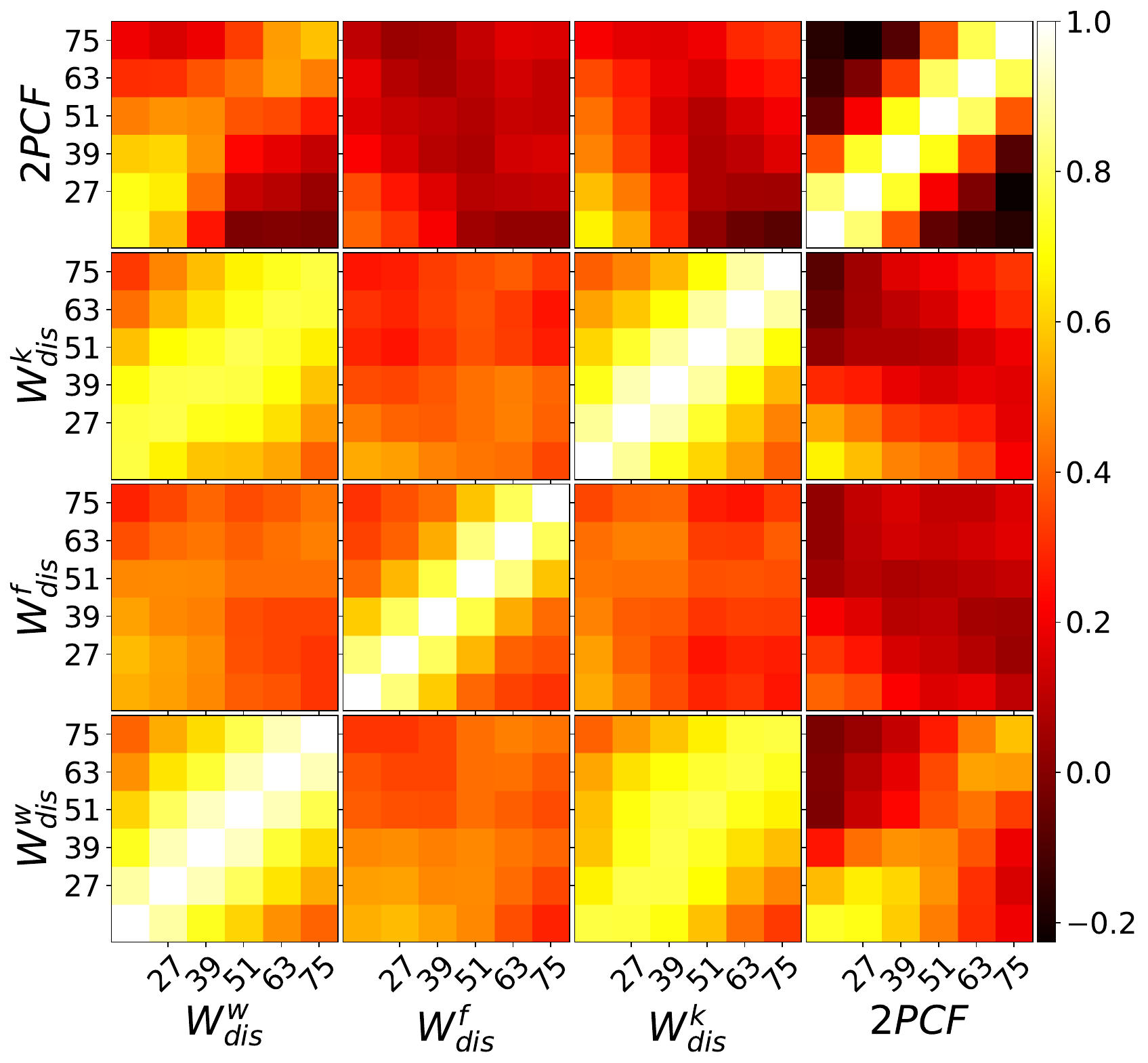"} 
    \includegraphics[width=0.47\textwidth]{"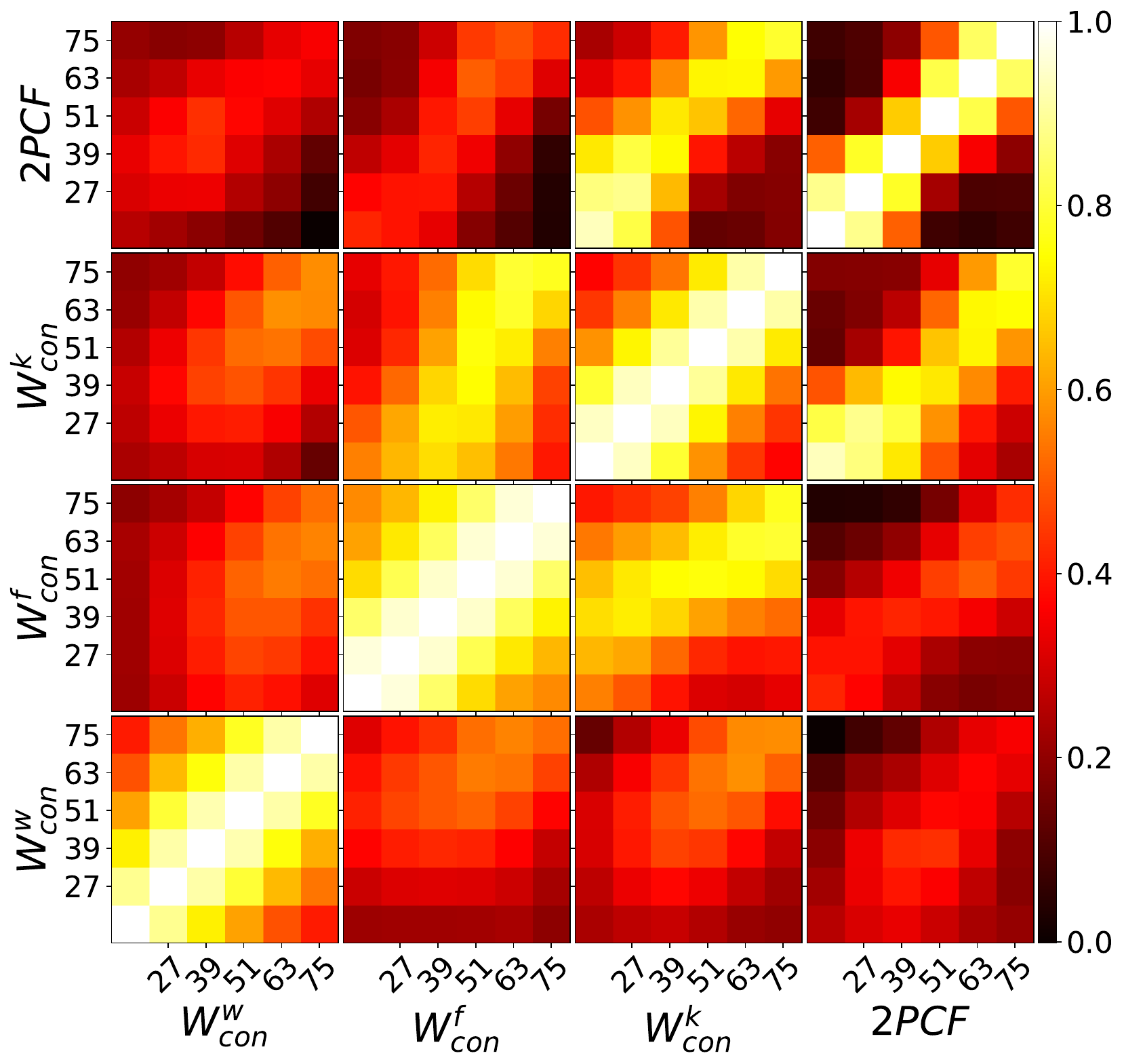"} 
\caption{Correlation matrices of the data covariance matrix $C_{\mathrm{data}}$ for the projected clustering statistics. The left panel shows the result for the discrete morphology marker, while the right panel shows the result for the continuous morphology strength marker. In each matrix, the data vector is ordered from lower left to upper right as the standard 2PCF, wall, filament, and knot statistics. The covariance is estimated using the coarse binning scheme with a fixed bin width of $10~h^{-1}{\rm Mpc}$.
}
\label{fig:correlation}
\end{figure*}

Figure~\ref{fig:correlation} shows the correlation coefficient matrices of the data covariance, defined as
$r_{ij} \equiv C_{ij}/\sqrt{C_{ii}C_{jj}}$. We use the correlation coefficient instead of the covariance itself because different morphology statistics can have very different amplitudes. This normalization makes it easier to compare the level of correlation among different parts of the data vector. The left panel shows the result for the discrete morphology marker, while the right panel shows the result for the continuous morphology strength marker. In both cases, the off-diagonal correlations between different morphology types are generally weaker than the correlations within the same type. This indicates that the knot, filament, and wall statistics are not fully redundant with each other. They therefore provide additional morphology-dependent clustering information beyond that contained in the standard 2PCF alone.

\subsection{Gaussian Process regression}

As described in Section~\ref{sec:data}, we apply three different mass cuts to each realization of the \textsc{Kun} simulation suite. These mass cuts correspond to three tracer samples with different bias values. For each cosmology and each mass cut, we compute the projected MCFs using both the discrete morphology marker and the continuous morphology strength marker defined from the \textsc{Nexus} classification. To include the dependence on tracer bias in the emulator, we introduce an additional parameter $b$, as defined in Equation~\ref{eq:bias}.

The input parameter space of the emulator therefore consists of eight cosmological parameters,
$\{\Omega_b, \Omega_m, n_s, H_0, A_s, w_0, w_a, \sum M_\nu\}$,
together with the tracer bias parameter $b$. For each cosmology, the three mass cuts provide three biased tracer samples, giving a total of $129 \times 3 = 387$ training data points. The emulator output is the projected marked correlation function $W(s)$.

We use Gaussian-process regression (GPR)~\citep{williams2006gaussian} to emulate the dependence of the marked correlation statistics on cosmological parameters. The goal is to obtain a fast and smooth prediction of the data vector at any cosmology within the training domain, without rerunning expensive simulations. For a cosmological parameter vector $\hat{\theta}$, we denote a MCF data vector by $w(\hat{\theta})$.  A Gaussian process places a prior distribution over functions,
\begin{equation}
f(\hat{\theta})
\sim
\mathcal{GP}
\left[
m(\hat{\theta}),
k(\hat{\theta},\hat{\theta}')
\right]\,,
\end{equation}
where $m(\hat{\theta})$ is the mean function and $k(\hat{\theta},\hat{\theta}')$ is the covariance kernel. The simulated statistic is modeled as
\begin{equation}
w(\hat{\theta}) = f(\hat{\theta}) + \epsilon,
\qquad
\epsilon \sim \mathcal{N}(0,\sigma_n^2),
\end{equation}
where $\sigma_n^2$ represents the emulator noise term.

Given training cosmologies $\hat{\theta}$ and a test cosmology $\hat{\theta}_*$, the joint distribution of the training values $f$ and the test value $f_*$ is
\begin{equation}
\begin{bmatrix}
f \\
f_*
\end{bmatrix}
\sim
\mathcal{N}
\left[
\begin{bmatrix}
m(\hat{\theta}) \\
m(\hat{\theta}_*)
\end{bmatrix},
\begin{bmatrix}
K(\hat{\theta},\hat{\theta})+\sigma_n^2 I
&
K(\hat{\theta},\hat{\theta}_*) \\
K(\hat{\theta}_*,\hat{\theta})
&
K(\hat{\theta}_*,\hat{\theta}_*)
\end{bmatrix}
\right]\,,
\end{equation}
where $K(\hat{\theta},\hat{\theta})$ is the kernel matrix evaluated over all training cosmologies.

Conditioning on the training simulations gives a Gaussian predictive distribution,
\begin{equation}
f_* \mid \mathcal{D}, \hat{\theta}_*
\sim
\mathcal{N}
\left(
\bar{f}_*,
{\rm Cov}(f_*)
\right)\,,
\end{equation}
where $\mathcal{D}=\{\hat{\theta},w\}$ denotes the training cosmologies and their simulated statistics. The predictive mean is
\begin{equation}
\bar{f}_*
=
m(\hat{\theta}_*)
+
K(\hat{\theta}_*,\hat{\theta})
\left[
K(\hat{\theta},\hat{\theta})+\sigma_n^2 I
\right]^{-1}
\left[
w-m(\hat{\theta})
\right]\,,
\end{equation}
and the predictive covariance is
\begin{equation}
{\rm Cov}(f_*)
=
K(\hat{\theta}_*,\hat{\theta}_*)
-
K(\hat{\theta}_*,\hat{\theta})
\left[
K(\hat{\theta},\hat{\theta})+\sigma_n^2 I
\right]^{-1}
K(\hat{\theta},\hat{\theta}_*)\,.
\end{equation}
The predictive mean $\bar{f}_*$ is used as the emulator prediction, while the predictive covariance quantifies the interpolation uncertainty.

The covariance is specified by the kernel function. The Gaussian process kernel is constructed as:
\begin{equation}
K(\hat{\theta}_i, \hat{\theta}_j) = C_{const} \cdot k_{\text{RBF}}(\hat{\theta}_i, \hat{\theta}_j)
\end{equation}
where
\begin{equation}
k_{\text{RBF}}(\hat{\theta}_i, \hat{\theta}_j) = \exp\left(-\frac{\|\hat{\theta}_i - \hat{\theta}_j\|^2}{2l^2}\right).
\end{equation}
where $l$ is the length scale.
where $C_{\rm const}$ represents the constant kernel that scales the overall covariance amplitude, and $k_{\text{RBF}}(\hat{\theta}_i, \hat{\theta}_j)$ denotes the base kernel.

The training process in GPR is to find the optimal hyperparameters for the given training data and kernel function. This is achieved by the optimization of the log marginal likelihood of the input data via maximizing

\begin{equation}
\begin{aligned}
\ln \mathcal{L} = & -\frac{1}{2} f^\top [K(\hat{\theta}, \hat{\theta}) + \sigma_n^2 I]^{-1} f \\
& - \frac{1}{2} \log |K(\hat{\theta}, \hat{\theta}) + \sigma_n^2 I| - \frac{n}{2} \log 2\pi.
\end{aligned}
\end{equation}

For the practical implementation of GPR, we utilize the Python library \textsc{scikit-learn} \cite{scikit-learn} in the training process.

To further evaluate the model's generalization performance, We employ the leave-one-out (LOO) cross-validation method which provides a nearly unbiased estimate of the prediction error for small datasets. Each simulation is utilized once as a validation run, while the remaining simulations form the training set. Thus, we can obtain the emulation errors for all training cosmologies. The LOO error is defined as the 68th percentile error (1$\sigma$) across all samples. 

\begin{figure*}[htpb]
	\centering
    \includegraphics[width=0.95\textwidth]{"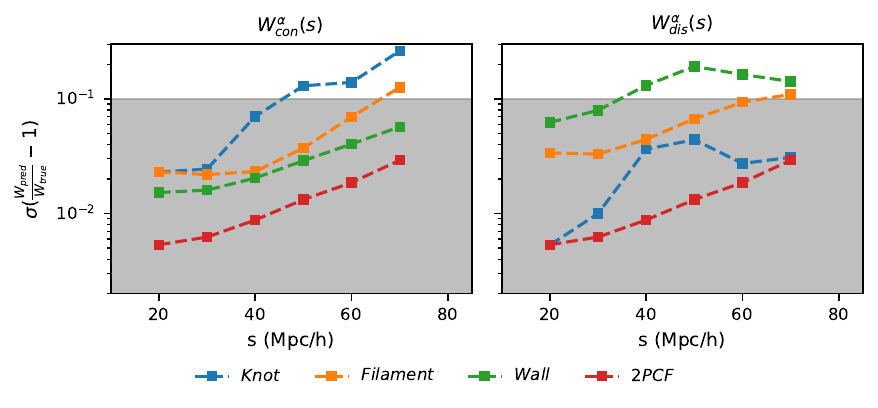"} 
\caption{The 68th-percentile fractional emulator errors from the leave-one-out validation. The left panel shows the result for the discrete morphology marker, while the right panel shows the result for the continuous morphology strength marker. Different colors correspond to the MCFs of different morphology types.}
\label{fig:LOO}
\end{figure*}

As shown in Figure~\ref{fig:LOO}, the leave-one-out validation indicates that the emulator performs reasonably well for most morphology statistics. For the continuous morphology strength marker, the relative prediction errors are within $13\%$ for all morphology types except knots. The larger errors for knots are expected, because knot regions are rare and highly clustered, and therefore suffer from stronger sample variance. For the discrete morphology marker, the errors are below $20\%$. Overall, the emulator accuracy is sufficient for the subsequent parameter inference. To account for this additional source of uncertainty, we include the emulator prediction error in the covariance used in the likelihood analysis.

\subsection{Covariance estimation}

With the GPR emulator, we can predict the marked-correlation data vector at any cosmology within the training domain. We now describe the covariance model and likelihood used to compare the emulator prediction with the measured data vector.

We denote the full data vector by $\mathbf{y}=W(s)$. For a cosmological parameter vector $\boldsymbol{\theta}$, the emulator prediction is denoted by $\mathbf{y}_{\rm emu}(\boldsymbol{\theta})$. Assuming a Gaussian likelihood, we write
\begin{equation}
\ln \mathcal{L}(\boldsymbol{\theta})
=
-\frac{1}{2}
\Delta\mathbf{y}^{\rm T}
\mathbf{C}^{-1}
\Delta\mathbf{y}
+ {\rm const.},
\label{eq:likelihood}
\end{equation}
where
\begin{equation}
\Delta\mathbf{y}
=
\mathbf{y}_{\rm obs}
-
\mathbf{y}_{\rm model}(\boldsymbol{\theta}).
\end{equation}
Here $\mathbf{y}_{\rm obs}$ is the measured data vector, and $\mathbf{y}_{\rm model}(\boldsymbol{\theta})$ is the phase-corrected emulator prediction defined below. Since the covariance matrix is fixed in our analysis, the normalization term of the Gaussian likelihood does not affect parameter inference and is absorbed into the constant.

The total covariance is modeled as
\begin{equation}
\mathbf{C}
=
\mathbf{C}_{\rm data}
+
\mathbf{C}_{\rm emu}\,.
\label{eq:total_cov}
\end{equation}
The first term describes the statistical uncertainty of the measured data vector, while the second term accounts for interpolation errors from the emulator. The effect of the fixed initial phase in the training simulations is treated as a correction to the emulator mean prediction, rather than as an additional covariance term.

\subsubsection{Data covariance}

We estimate the data covariance from subvolumes of the \textsc{Jiutian} simulation. The full simulation is divided into $N_{\rm sub}=5^3=125$ subsamples, each with volume
 $V_{\rm sub} = (400~h^{-1}{\rm Mpc})^3$. Let $\mathbf{y}_k$ be the data vector measured from the $k$-th subvolume, and let
\begin{equation}
\bar{\mathbf{y}}
=
\frac{1}{N_{\rm sub}}
\sum_{k=1}^{N_{\rm sub}}
\mathbf{y}_k
\end{equation}
be the mean over all subvolumes. The covariance for an observed volume $V_{\rm obs}$ is then estimated as
\begin{equation}
\mathbf{C}_{\rm data}
=
\frac{V_{\rm sub}}{V_{\rm obs}}
\frac{1}{N_{\rm sub}-1}
\sum_{k=1}^{N_{\rm sub}}
\left(
\mathbf{y}_k-\bar{\mathbf{y}}
\right)
\left(
\mathbf{y}_k-\bar{\mathbf{y}}
\right)^{\rm T}.
\label{eq:data_cov}
\end{equation}
The factor $V_{\rm sub}/V_{\rm obs}$ rescales the covariance from the subvolume size to the target observational volume.

\subsubsection{Emulator covariance}

The emulator covariance quantifies the uncertainty introduced by interpolation across cosmological parameter space. We estimate this term using the leave-one-out validation described in the previous section. For the $j$-th training cosmology, the emulator is trained on all other cosmologies and then evaluated at the omitted point. The residual vector is
\begin{equation}
\mathbf{e}_j
=
\mathbf{y}^{(-j)}_{\rm emu}(\boldsymbol{\theta}_j)
-
\mathbf{y}_{\rm true}(\boldsymbol{\theta}_j),
\end{equation}
where $\mathbf{y}^{(-j)}_{\rm emu}$ is the leave-one-out emulator prediction and $\mathbf{y}_{\rm true}$ is the statistic measured directly from the simulation.

The emulator covariance is estimated from these residuals:
\begin{equation}
\mathbf{C}_{\rm emu}
=
\frac{1}{N_{\rm train}-1}
\sum_{j=1}^{N_{\rm train}}
\left(
\mathbf{e}_j-\bar{\mathbf{e}}
\right)
\left(
\mathbf{e}_j-\bar{\mathbf{e}}
\right)^{\rm T},
\label{eq:emu_cov}
\end{equation}
where
\begin{equation}
\bar{\mathbf{e}}
=
\frac{1}{N_{\rm train}}
\sum_{j=1}^{N_{\rm train}}
\mathbf{e}_j .
\end{equation}
In practice, we find that $\mathbf{C}_{\rm emu}$ is much smaller than $\mathbf{C}_{\rm data}$, indicating that emulator interpolation errors are subdominant compared with the statistical uncertainty of the data vector.

\subsubsection{Correction for fixed initial phases}

The \textsc{Kun} training simulations share fixed initial phases. This reduces simulation noise in the emulator training set, but it can also shift the emulator prediction away from the ensemble-averaged statistic expected from independent realizations. We correct this effect at the level of the emulator mean prediction, following the ratio-based approach of \cite{Yuan_2022}.

We define a bin-by-bin phase-correction factor,
\begin{equation}
R_i=\frac{
\bar{y}_{{\rm Jiutian},i}
}{
y_{{\rm Kun,fid},i}
}\,.
\end{equation}
The corrected emulator prediction is then
\begin{equation}
y_{{\rm model},i}(\boldsymbol{\theta})=R_i y_{{\rm emu},i}(\boldsymbol{\theta})\,,
\label{eq:phase_correction}
\end{equation}
where $i$ labels the elements of the data vector.

This correction assumes that the fixed-phase offset is weakly dependent on cosmology. The ratio in Equation~\ref{eq:phase_correction} therefore maps the emulator prediction from the fixed-phase \textsc{Kun} realization to the ensemble mean estimated from the independent \textsc{Jiutian} realizations. Since the residual phase uncertainty is subdominant for our data vector, we use this corrected emulator mean in the fiducial likelihood and do not add a separate phase-covariance term.

\subsubsection{Inverse covariance correction}

Because the covariance matrix is estimated from a finite number of subvolumes, its inverse is biased. We correct this bias using the Hartlap factor~\citep{Hartlap_2006}. The debiased inverse covariance is
\begin{equation}
\widehat{\mathbf{C}}^{-1}
=
\frac{
N_{\rm sub}-N_d-2
}{
N_{\rm sub}-1
}
\mathbf{C}^{-1},
\label{eq:hartlap}
\end{equation}
where $N_d$ is the dimension of the data vector. This correction is well defined only when $N_{\rm sub}>N_d+2$, which motivates the compact binning scheme adopted for the MCF data vector.

\section{Results}
\label{sec:results}
\subsection{Comparison of 2PCF and MCFs}

In this section, we compare the cosmological constraining power of the standard 2PCF and the morphology-based MCFs. The measurement is performed on a mock halo catalog constructed by randomly selecting an $800~({\rm Mpc}/h)^3$ volume from the \textsc{Jiutian} simulation.

We run an MCMC analysis with the \textsc{emcee} package~\citep{emcee}. The data vector uses separations in the range $15~h^{-1}{\rm Mpc} \leq s \leq 75~h^{-1}{\rm Mpc}$, with a bin width of $\Delta s = 10~h^{-1}{\rm Mpc}$. We adopt a halo mass threshold of $M_{\rm cut}=3.85\times10^{12}~h^{-1}M_\odot$. The free parameters are $\Omega_m$, $\sigma_8$, and the halo bias parameter $b$. We use uniform priors with $\Omega_m \in [0.20,0.40]$, $\sigma_8 \in [0.54,1.18]$, and $b \in [1.0,3.0]$.

\begin{figure*}[htpb]
	\centering
	\begin{tabular}{cc}
		\includegraphics[width=0.48\textwidth]{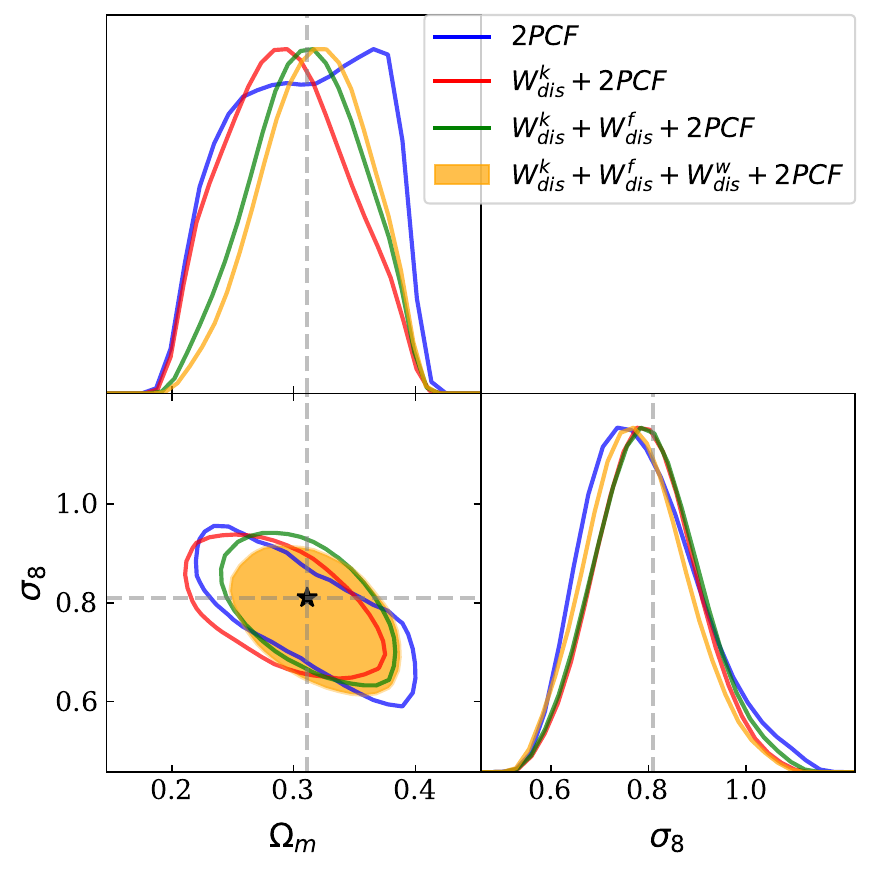} & 
		\includegraphics[width=0.48\textwidth]{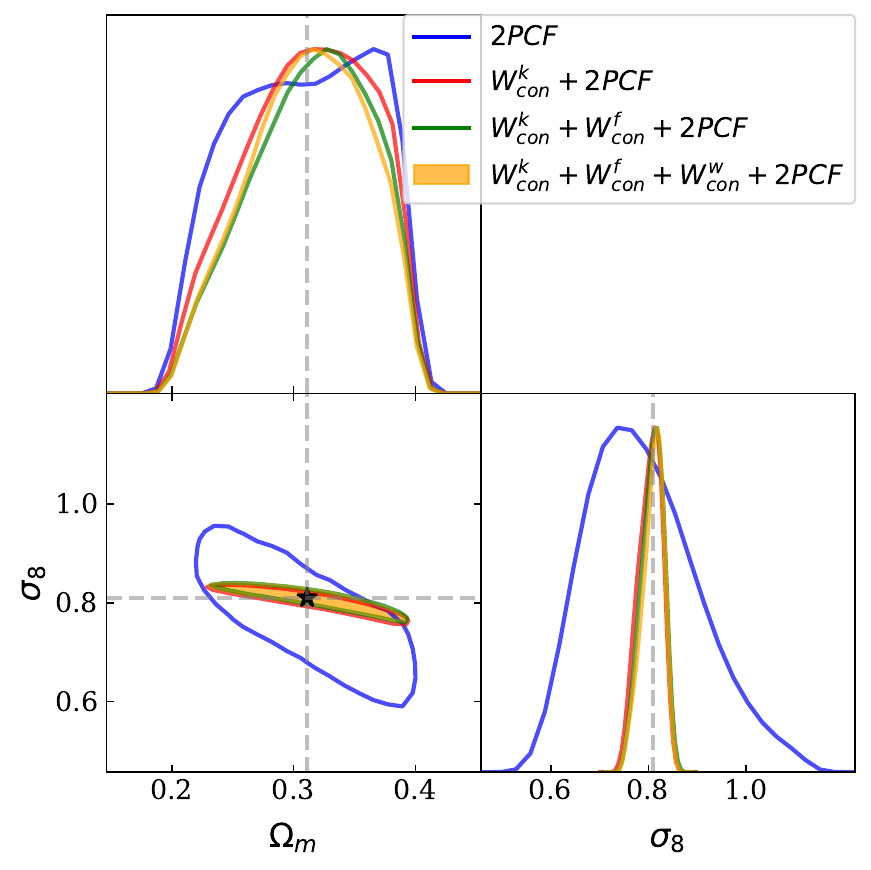} \\
	\end{tabular}
\caption{Posterior constraints on $\Omega_m$ and $\sigma_8$ at the $1\sigma$ confidence level. The left panel shows the results for the discrete morphology marker, while the right panel shows the results for the continuous morphology strength marker. In each panel, we compare the constraints from the standard 2PCF with those from different MCF data-vector combinations. The fiducial values, $\Omega_m=0.31$ and $\sigma_8=0.81$, are indicated by the dashed lines.}
\label{fig:mcmc_results_keys}
\end{figure*}

Figure~\ref{fig:mcmc_results_keys} shows the posterior constraints on $\Omega_m$ and $\sigma_8$, with the halo bias parameter $b$ marginalized over. The left panel corresponds to the discrete morphology marker, and the right panel corresponds to the continuous morphology strength marker. The fiducial cosmology is indicated by the black star. In each panel, we compare the standard 2PCF with the joint MCF analysis. The fiducial cosmology, $\Omega_m=0.31$ and $\sigma_8=0.81$, is marked by the black star.

For the standard 2PCF alone, we obtain $\Omega_m = 0.310^{+0.061}_{-0.065}$ and $\sigma_8 = 0.779^{+0.128}_{-0.104}$. Adding the discrete morphology MCFs gives $\Omega_m = 0.318^{+0.043}_{-0.046}$ and $\sigma_8 = 0.774^{+0.106}_{-0.097}$. Thus, the discrete marker improves the constraint on $\Omega_m$ by about $30\%$, but gives little improvement on $\sigma_8$.

The continuous morphology strength marker gives a much stronger improvement. When combined with the 2PCF, it yields
$\Omega_m = 0.314^{+0.049}_{-0.053}$ and
$\sigma_8 = 0.811^{+0.019}_{-0.026}$. Compared with the 2PCF-only result, the uncertainty on $\sigma_8$ is reduced by roughly a factor of five, while the improvement on $\Omega_m$ remains modest. This shows that the continuous strength marker captures additional morphology-dependent clustering information and is particularly effective at tightening the constraint on the fluctuation amplitude.

\begin{figure}[htpb]
	\centering
	\includegraphics[width=0.48\textwidth]{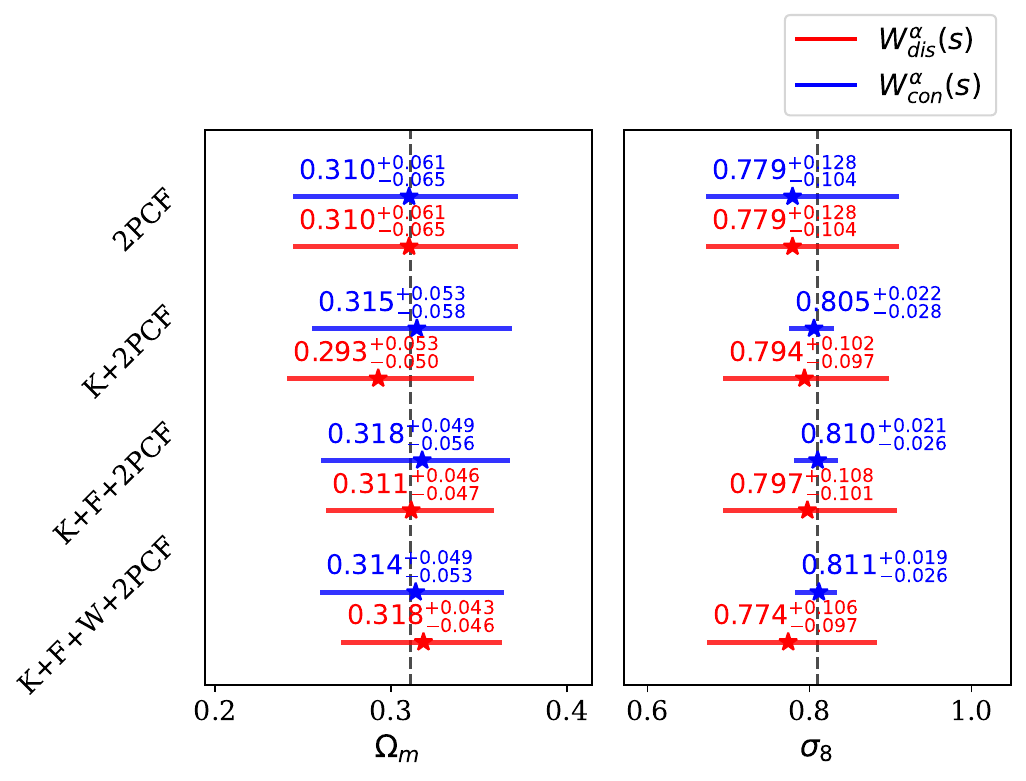} 
\caption{
One-dimensional marginalized posterior distributions of $\Omega_m$ and $\sigma_8$ from the standard 2PCF and different morphology-marker configurations. The fiducial values, $\Omega_m=0.31$ and $\sigma_8=0.81$, are indicated by the dashed lines.
}
\label{fig:mcmc_results_keys_1d}
\end{figure}

To compare the constraints more clearly, Figure~\ref{fig:mcmc_results_keys_1d} shows the one-dimensional marginalized posterior distributions of $\Omega_m$ and $\sigma_8$, produced with the \textsc{getdist} package~\citep{Lewis_2025}. This figure summarizes the results from both morphology-marker choices. The continuous morphology strength marker gives a much tighter constraint on $\sigma_8$, while the discrete morphology marker provides only a modest improvement in $\Omega_m$ and little improvement in $\sigma_8$.

To compare the overall constraining power of different data-vector combinations, we use the Figure of Merit (FoM), defined as \citep{Wang2008FOM}
\begin{equation}
\mathrm{FoM} \equiv \frac{1}{\sigma(\Omega_m)\sigma(\sigma_8)\sqrt{1-\rho^2}},
\label{eq:FoM}
\end{equation}
where $\rho$ is the correlation coefficient between $\Omega_m$ and $\sigma_8$. A larger FoM corresponds to a tighter joint constraint, as it accounts for both the marginalized parameter uncertainties and their degeneracy.

To facilitate comparison between different methods, we further define the FoM ratio,
\begin{equation}
R_{\rm FoM} \equiv \frac{\rm FoM}{\rm FoM(\rm 2PCF)}\,,
\label{eq:RFoM}
\end{equation}
where FoM$(\rm 2PCF)$ is the FoM obtained using the 2PCF alone. The FoM ratio therefore quantifies the improvement in constraining power relative to the 2PCF, with $R_{\rm FoM}>1$ indicating tighter constraints.

\begin{table}[htbp]
\centering
\caption{
$R_{\rm FoM}$ for different data-vector combinations. For each marker scheme, the FoM is normalized by the result from the full combination of the 2PCF and all three morphology MCFs.
}
\label{tab:fom_keys}
\begin{tabular}{lcc}
\hline
Combination & Continuous    &  Discrete \\
\hline
2PCF only     & 1.000 & 1.000 \\
$+$ knot      & 7.820 & 1.107 \\
$+$ knot + filament  & 8.454 & 1.141 \\
$+$ knot + filament + wall & 8.618 & 1.176 \\
\hline
\end{tabular}
\end{table}

$R_{\rm FoM}$ in Table~\ref{tab:fom_keys} show a clear difference between the two morphology-marker choices. For the continuous morphology strength marker, most of the constraining power is already obtained from the knot-weighted MCF. $R_{\rm FoM}$ increases from unity for the 2PCF alone to $7.820$ when the knot strength is added, corresponding to an improvement by a factor of about eight. Adding the filament and wall strength markers only gives small further increases, and the FoM reaches $8.618$ for the full data-vector combination by construction. This indicates that the strength-based MCFs gain most of their cosmological information from dense knot regions, while filaments and walls provide secondary contributions.

For the discrete morphology marker, the improvement is much weaker. $R_{\rm FoM}$ increases only gradually, from unity for the 2PCF alone to $1.176$ when all three morphology-selected MCFs are included. This corresponds to an improvement of less than $20\%$. No single morphology type dominates the information gain, and the overall improvement over the standard 2PCF remains limited.

Overall, the constraining power of the MCFs depends strongly on how the morphology marker is defined. The continuous strength marker extracts substantially more information than the discrete marker, improving the FoM by nearly an order of magnitude and reducing the uncertainty on $\sigma_8$ by about a factor of five. The dominant contribution comes from knot regions, suggesting that dense cosmic-web environments carry particularly strong non-Gaussian information. In contrast, the discrete marker provides only modest additional constraining power and does not significantly change the $\Omega_m$--$\sigma_8$ degeneracy.

\subsection{Constraints from Different MCF Configurations}

We next examine how the smoothing scheme and the separation range affect the cosmological constraints from MCFs. We consider the combined data vectors formed by the standard 2PCF and all three morphology MCFs, separately for the discrete morphology marker and the continuous morphology strength marker. The constraining power of each configuration is quantified by the FoM ratio defined in Equation~\ref{eq:RFoM}.
\begin{figure*}[htpb]
    \centering
    \begin{tabular}{cc}
        \includegraphics[width=0.48\textwidth]{"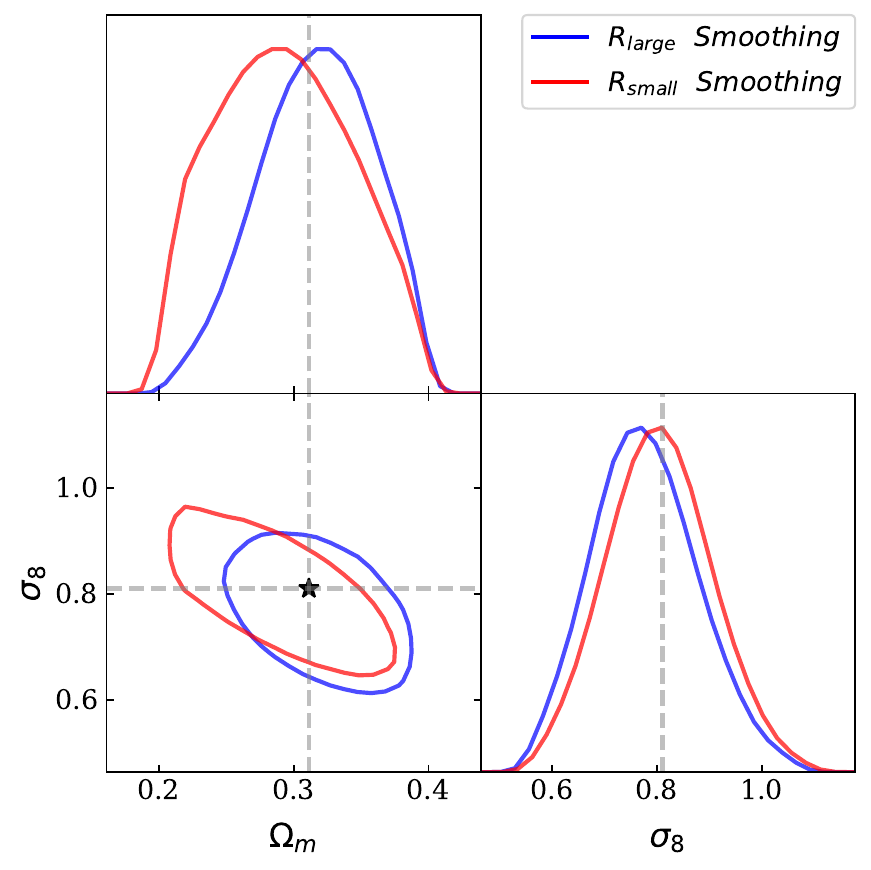"} & \includegraphics[width=0.48\textwidth]{"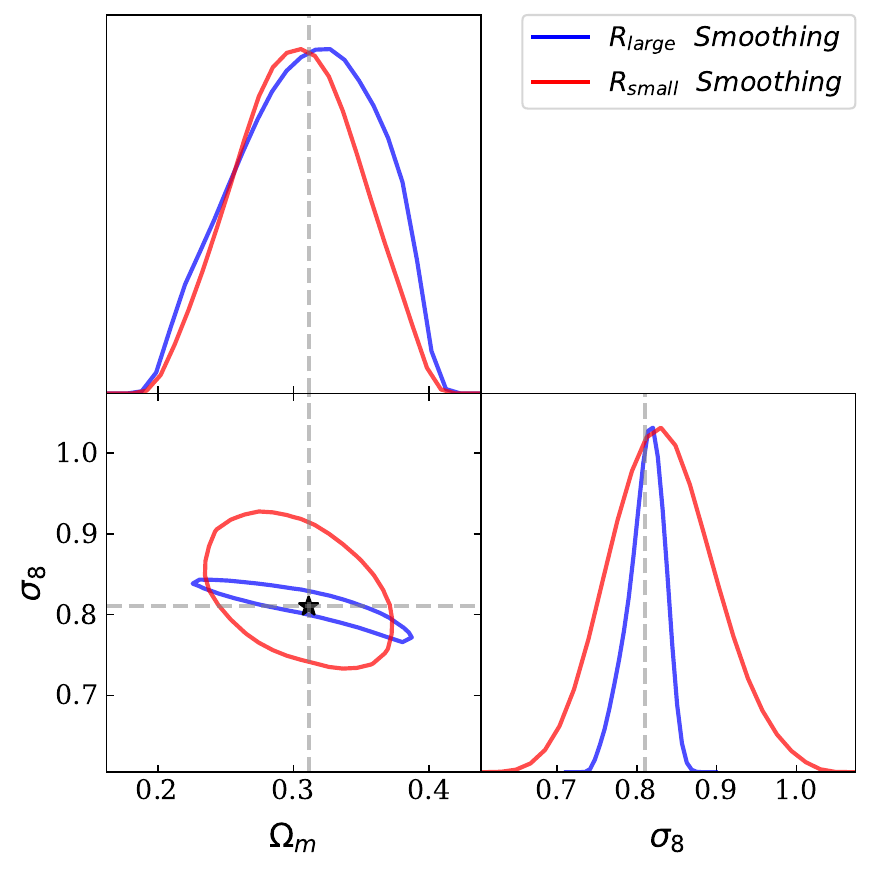"} \\
    \end{tabular}
    \caption{
MCMC constraints for different smoothing schemes. The left panel shows the results for the discrete morphology marker, while the right panel shows the results for the continuous morphology strength marker. The contours indicate the 68\% credible regions in the $\Omega_m$--$\sigma_8$ plane.
}
    \label{fig:mcmc_smoothing_scales}
\end{figure*}

\begin{table}[htbp]
\centering
\caption{
$R_{\rm FoM}$  for different smoothing schemes. All values are normalized by the fiducial configuration.}
\label{tab:fom_smoothing}
\begin{tabular}{lcc}
\hline
Smoothing scheme & Continuous  & Discrete \\
\hline
$R_{\rm large}$ & $8.618$ & $1.176$ \\
$R_{\rm small}$ & $1.779$ & $1.143$ \\
\hline
\end{tabular}
\end{table}

We first test the impact of the smoothing scheme. As shown in Table~\ref{tab:fom_smoothing} and Figure~\ref{fig:mcmc_smoothing_scales}, the continuous morphology strength marker is strongly affected by this choice. The large-scale smoothing scheme gives $R_{\rm FoM}=8.618$, while the small-scale smoothing scheme reduces $R_{\rm FoM}$ to $1.779$. This indicates that the continuous strength marker is most informative when the morphology is defined from a smoothed large-scale density field. In contrast, the discrete morphology marker is nearly insensitive to the smoothing scheme, with $R_{\rm FoM}$ changing only from $1.176$ to $1.143$.

\begin{figure*}[htpb]
    \centering
    \begin{tabular}{cc}
        \includegraphics[width=0.48\textwidth]{"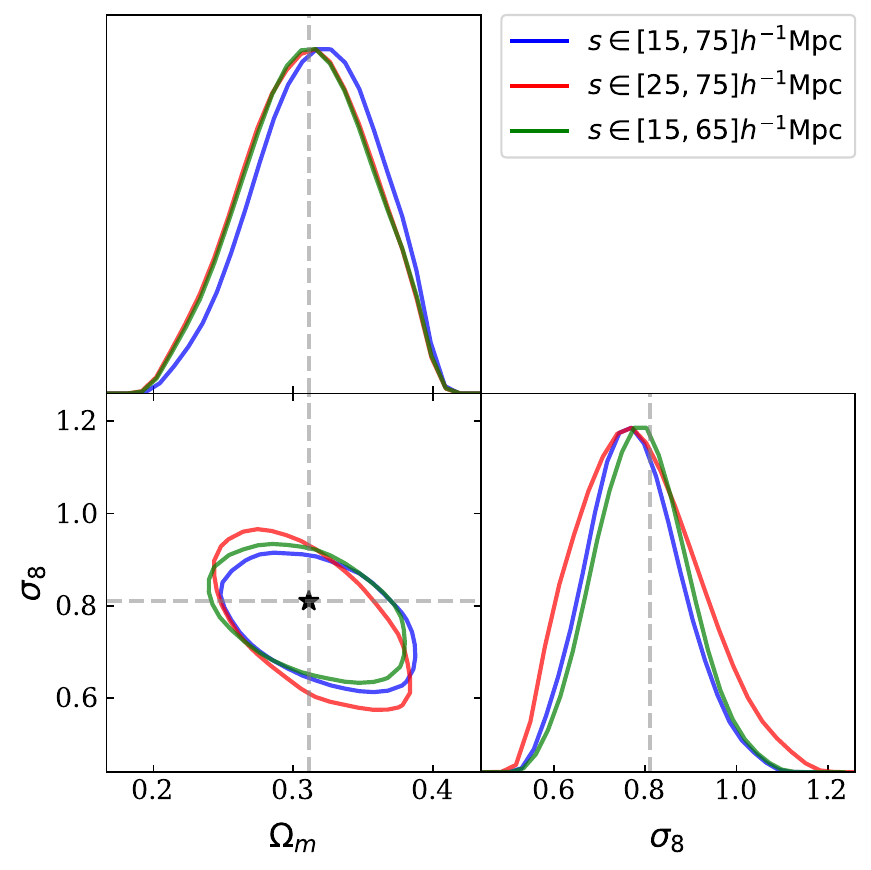"} & \includegraphics[width=0.48\textwidth]{"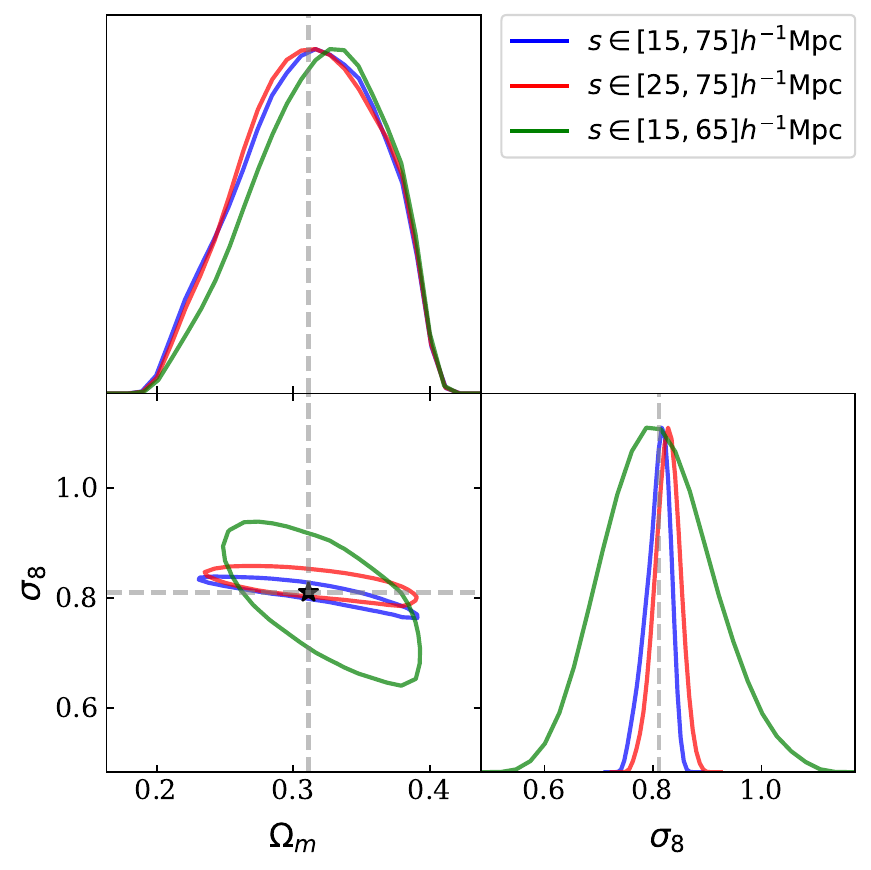"} \\
    \end{tabular}
\caption{MCMC constraints for different separation ranges, $s \in [25,75]$, $[15,75]$, and $[15,65]~h^{-1}{\rm Mpc}$. The left panel shows the results for the discrete morphology marker, while the right panel shows the results for the continuous morphology strength marker. The contours indicate the 68\% credible regions in the $\Omega_m$--$\sigma_8$ plane.}
    \label{fig:mcmc_s_range}
\end{figure*}

\begin{table}[htbp]
\centering
\caption{
$R_{\rm FoM}$ for different separation ranges. All values are normalized by the fiducial configuration.
}
\label{tab:fom_srange}
\begin{tabular}{lcc}
\hline
$s$ ($h^{-1}{\rm Mpc}$) & Continuous & Discrete \\
\hline
$[25,75]$ & $5.312$ & $0.961$ \\
$[15,75]$ & $8.618$ & $1.176$ \\
$[15,65]$ & $1.301$ & $1.152$ \\
\hline
\end{tabular}
\end{table}

We then examine the dependence on the separation range. For the continuous morphology strength marker, Table~\ref{tab:fom_srange} and Figure~\ref{fig:mcmc_s_range} show that both ends of the scale range contribute to the final constraint. Removing the smallest bin, by using $s \in [25,75]\,h^{-1}{\rm Mpc}$, lowers $R_{\rm FoM}$ from $8.618$ to $5.312$. Removing the largest bin, by using $s \in [15,65]\,h^{-1}{\rm Mpc}$, has an even stronger effect and reduces $R_{\rm FoM}$ to $1.301$. Therefore, the continuous strength marker requires the full separation range to reach its best performance.

The discrete morphology marker shows a much weaker scale dependence. $R_{\rm FoM}$ decreases from $1.176$ to $0.961$ when the smallest scales are removed, but remains almost unchanged when the largest scales are removed. This suggests that the discrete marker gains some information from the $s\in [15, 25]~h^{-1}{\rm Mpc}$ bin, while the $s\in [65, 75]~h^{-1}{\rm Mpc}$ bin contributes little.

Overall, the continuous morphology strength marker provides much stronger constraints, but its performance depends sensitively on the smoothing scheme and the adopted separation range. The discrete morphology marker gives weaker constraints, but is more stable under these analysis choices.

\subsection{Sensitivity to halo mass cut}

To test the sensitivity of our constraints to the halo selection, we vary the mass threshold in the mock data while keeping the emulator training set unchanged. We consider three thresholds,
$M_{\rm cut}=\{3.85, 2.45, 0.85\}\times10^{12}~h^{-1}M_{\odot}$.
Changing $M_{\rm cut}$ changes the typical host halo mass and therefore the effective tracer bias. For each threshold, we randomly subsample the halo catalog to the same number density. This removes the trivial effect of different tracer abundances and keeps the shot noise approximately fixed, so that the comparison mainly reflects the impact of halo mass selection and the associated change in tracer bias.

\begin{figure*}[htpb]
\centering
\begin{tabular}{cc}
\includegraphics[width=0.48\textwidth]{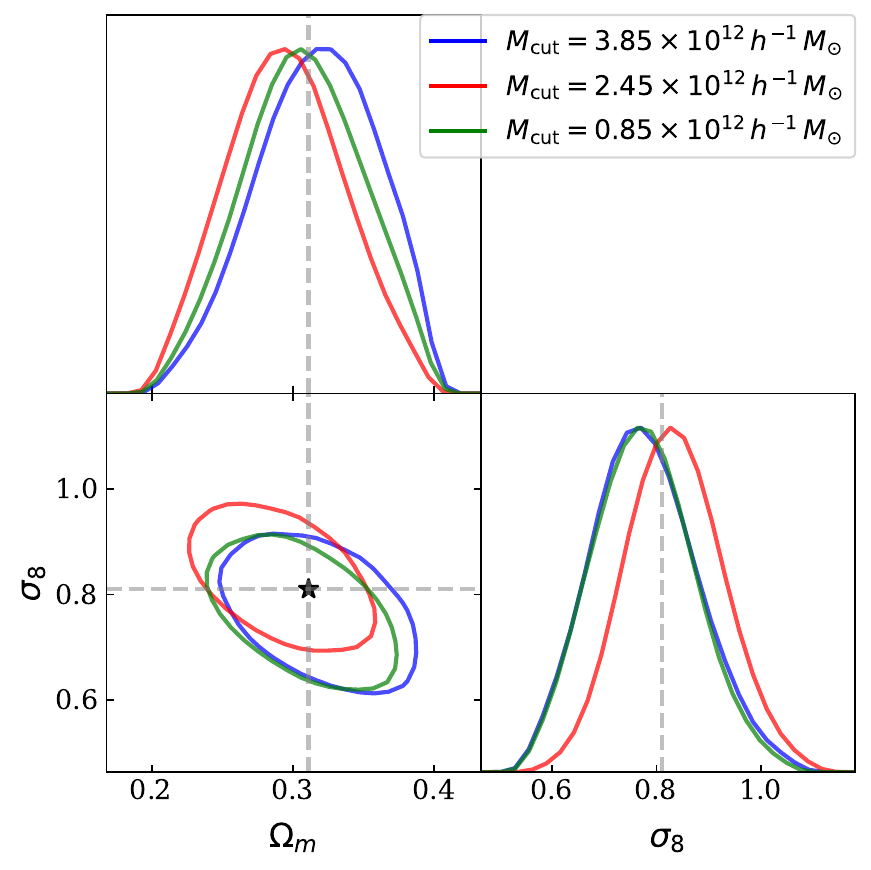} &
\includegraphics[width=0.48\textwidth]{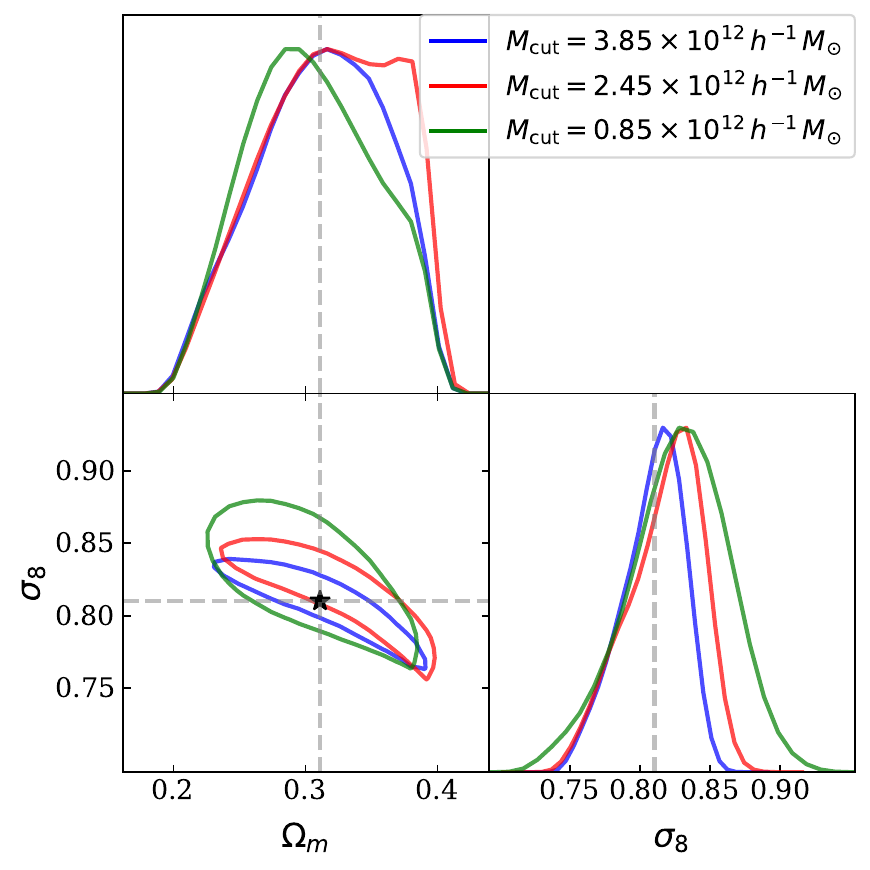} \
\end{tabular}
\caption{
MCMC constraints for different halo mass thresholds:
$M_{\rm cut}=3.85\times10^{12}$,
$2.45\times10^{12}$, and
$0.85\times10^{12}~h^{-1}M_{\odot}$. The left panel shows the results for the discrete morphology marker, while the right panel shows the results for the continuous morphology strength marker. In all cases, the number density of tracers is fixed, so that the comparison mainly reflects the impact of halo mass selection rather than changes in tracer abundance.
}
\label{fig:mcmc_results_mass_shift}
\end{figure*}

The results are summarized in Table~\ref{tab:mass_shift_fom} and Figure~\ref{fig:mcmc_results_mass_shift}. For the discrete morphology marker, $R_{\rm FoM}$ remains low and depends only weakly on the halo mass threshold. It changes from $1.176$ for the high-mass cut to $1.322$ for the intermediate-mass cut and $1.257$ for the low-mass cut. The posterior contours remain consistent with the fiducial cosmology for all three mass cuts. This indicates that the discrete marker is relatively insensitive to the halo mass selection, but its overall constraining power is limited.

For the continuous morphology strength marker, the constraining power decreases as the mass threshold is lowered. $R_{\rm FoM}$ drops from $8.618$ for the high-mass cut to $6.012$ for the intermediate-mass cut and $3.358$ for the low-mass cut. This trend suggests that the strength marker is more sensitive to the tracer population. Higher-mass halos have a larger effective bias and trace the large-scale density field more strongly, allowing the strength-weighted MCF to extract more cosmological information. For lower-mass halos, the tracer bias is smaller, and the gain from the strength marker is reduced. Even for the lowest mass threshold, the continuous strength marker still gives $R_{\rm FoM}= 3.358$, which is substantially higher than the 2PCF-only value of unity. In addition, the posterior contours remain consistent with the fiducial values within the 68\% credible regions.

Therefore, the continuous strength marker remains unbiased over the tested mass range, although its precision degrades for lower-mass halo samples. These results suggest that the continuous morphology strength marker can give stronger constraints, but is more sensitive to halo selection and tracer bias. The discrete morphology marker gives weaker constraints, but is more stable across different mass thresholds.

\begin{table}[htbp]
\centering
\caption{
$R_{\rm FoM}$ for different halo mass thresholds. All values are normalized by the fiducial configuration.
}
\label{tab:mass_shift_fom}
\begin{tabular}{lcc}
\hline
$M_{\rm cut}~(\times 10^{12}~h^{-1}M_{\odot})$ & Continuous  & Discrete  \\
\hline
$3.85$ & $8.618$ & $1.176$ \\
$2.45$ & $6.012$ & $1.322$ \\
$0.85$ & $3.358$ & $1.257$ \\
\hline
\end{tabular}
\end{table}

\section{Conclusion}
\label{sec:conclusion}

In this work, we introduced morphology-based MCFs for halo clustering. The morphology markers are derived from the \textsc{Nexus} cosmic-web classification and its associated strength fields. We considered two choices of marker: a discrete morphology marker, which separates halos into knots, filaments, and walls, and a continuous morphology strength marker, which weights halos by their local \textsc{Nexus} strength.

To use these statistics for cosmological inference, we built a Gaussian-process emulator trained on the \textsc{Kun} simulation suite. The emulator includes eight cosmological parameters and an additional tracer bias parameter. Leave-one-out tests show that the emulator reproduces the projected MCFs with sufficient accuracy for the likelihood analysis, and the emulator uncertainty is included in the covariance.

Using mock halo catalogues from the \textsc{Jiutian} simulation, we compared the constraining power of the standard 2PCF and the morphology-based MCFs. The continuous morphology strength marker provides the strongest improvement. When combined with the 2PCF, it increases $R_{\rm FoM}$ by a factor of about $8.6$ relative to the 2PCF alone and reduces the uncertainty on $\sigma_8$ by roughly a factor of five. Most of this gain comes from the knot strength, indicating that dense cosmic-web regions carry particularly strong cosmological information. In contrast, the discrete morphology marker gives only a modest improvement, increasing the FoM by less than $20\%$.

We also tested the dependence on smoothing scale, separation range, and halo mass threshold. The continuous strength marker is most effective when the morphology field is defined using large-scale smoothing and when the full separation range is included. Its constraining power decreases when either the smallest or largest separation bin is removed. The discrete marker is less sensitive to these analysis choices, but its overall constraining power is much weaker.

The halo-mass tests show that the continuous strength marker is more sensitive to the tracer population and the associated effective bias. Its FoM decreases as the mass threshold is lowered, but the inferred parameters remain consistent with the fiducial cosmology for all mass cuts tested. Even for the lowest mass threshold, the continuous strength marker still outperforms the 2PCF alone. The discrete marker is more stable across mass thresholds, although it provides weaker constraints.

Overall, our results suggest that morphology-based MCFs capture additional clustering information that is not fully encoded in the standard 2PCF. The continuous morphology strength marker is the more powerful statistic, while the discrete morphology marker provides a more conservative and stable alternative. This makes morphology-based MCFs a promising probe for future applications to large-scale structure surveys.

\section{Acknowledgement}
This work is supported by National SKA Program of China (2025SKA0160100), National Science Foundation of China (12473097), the China Manned Space Project with No. CMS-CSST-2021 (A02, A03, B01),  Guangdong Basic and Applied Basic Research Foundation (2024A1515012309), the National Natural Science Foundation of China (12373005). We also acknowledge the Beijing Super Cloud Center (BSCC) and Beijing Beilong Super Cloud Computing Co., Ltd (\url{http://www.blsc.cn/}) for providing HPC resources that have significantly contributed to the research results presented in this paper.

\bibliography{apssamp}

\end{document}